\documentclass[12pt]{article}

\usepackage[fleqn]{amsmath}
\usepackage{subfig}
\usepackage{chemformula}

\title{Numerical simulation of ionospheric depletions resulting from rocket launches using a general circulation model}

\author{G. W. Bowden, P. Lorrain, and M. Brown \\
School of Engineering and Information Technology \\
University of New South Wales Canberra \\
Northcott Drive, Campbell \\
 ACT \emph{2612}, Australia \\
\texttt{g.bowden@adfa.edu.au}
}

\begin{document}
\maketitle

\begin{abstract}
Rocket exhaust plumes have been observed to cause large-scale depletions of ionospheric plasmas (``ionospheric holes''). In the F-region, charge exchange reactions occur between \ch{O^+} ions and exhaust species such as \ch{H_2O}, \ch{H_2}, and \ch{CO_2}  to form ions which then undergo rapid dissociative recombination. The Global Ionosphere-Thermosphere Model (GITM) was extended to include these chemical reactions and appropriate source terms to represent rocket exhaust plumes. The resulting model was applied to ionospheric depletions resulting from the launches of Jason-3 and FORMOSAT-5 on SpaceX Falcon 9 rockets from Vandenberg Air Force Base. Outputs from the model were compared with GNSS, ionosonde, and satellite Langmuir probe measurements. Simulation indicated that the FORMOSAT-5 launch resulted in a far larger and longer-lived ionospheric depletion than the Jason-3 launch, consistent with observations.
\end{abstract}

\section{Introduction}





The ionosphere represents the region of the Earth's upper atmosphere where significant quantities of ions and electrons occur. Rocket launches can result in significant perturbations to this plasma, including chemical depletion of the \ch{O^+} dominated F-region by rocket exhaust gases. Such ionospheric depletions may impact high frequency radio communications or Global Navigation Satellite System (GNSS) signals. These depletions also provide a test to help validate numerical models of the ionosphere-thermosphere system.

A range of measurement techniques have been used to observe the depletion of the ionosphere by rocket exhaust since the dawn of the space age. The  detection of an ionospheric hole was first reported by Booker \cite{booker1961} using ionosonde measurements following a Vanguard II missile launch in February 1959. Subsequently, such depletions have been studied using geostationary satellite signal Faraday rotation \cite{mendillo1975}, airglow emission \cite{mendillo1982}, incoherent scatter radar \cite{wand1984}, satellite Langmuir probe \cite{park2016}, and GNSS total electron content (TEC) \cite{furuya2008,mendillo2008,chou2018} measurements. Chou et al. \cite{chou2018} reported GPS observations of a large-scale ionospheric depletion resulting from the launch of FORMOSAT-5.

Analytical and numerical techniques have previously been applied to model these ionospheric depletions. Bernhardt, Park, and Banks \cite{bernhardt1975} derived an analytical expression for diffusion of rocket exhaust gases from the continuity equation and considered their injection into a numerical magnetic flux tube model. Subsequently, Bernhardt \cite{bernhardt1979} developed a numerical model for this process incorporating chemical reactions, diffusion of multiple species, and neutral winds. Mendillo \cite{mendillo1978} analytically derived expressions for the diffusion, chemical depletion, and recovery processes assuming the first two processes proceeded much faster than the third and neglecting changes in the ambient atmosphere with time. 

In this work, a numerical model is adapted to simulate ionospheric depletions due to rocket launches and its output is compared to measurements for two launches. An overview of the launches considered is provided in section~\ref{sec_F9_launches}. The numerical model used and the ways in which it is adapted to simulate ionospheric depletions are described in section~\ref{sec_modelling}. Results of the numerical simulation are presented in section~\ref{sec_results} and compared with observational data in section~\ref{sec_observation}. These results are further discussed and possible extensions to this work identified in section~\ref{sec_discussion}.


\section{Falcon 9 launches} \label{sec_F9_launches}


This study investigated depletion of the ionosphere due to launches of the Jason-3 and FORMOSAT-5 satellites. These launches took place at 18:42:18 UT 17 January 2016 and at 18:51:00 UT 24 August 2017 respectively from Vandenberg Air Force Base (located at $34^{\circ}44' \, \mathrm{N}$, $120^{\circ}34' \, \mathrm{W}$). In both cases SpaceX Falcon 9 launch vehicles were used (v1.1 in the former case and v1.2/Full Thrust in the latter). The depletions were attributed to the second stage Merlin 1D Vacuum engines, which operated at F-region heights using liquid oxygen (LOX) and kerosene (RP-1) propellants. Specifications for the Falcon 9 second stage are listed in Table~\ref{tab_F9_data}. Complete combustion of LOX and RP-1 produces \ch{H_2O} and \ch{CO_2}, while exhaust from incomplete combustion also contains \ch{H_2} and \ch{CO}.

\begin{table}
\caption{Falcon 9 second stage data from versions of the SpaceX website (https://www.spacex.com/falcon9) archived on 29 November 2013 and 9 December 2015.}
\label{tab_F9_data}
\begin{tabular}{@{}lcc}
 \hline
 & v1.1 & v1.2 \\
 \hline
 Burn time, $\Delta T$ & $375 \, \mathrm{s}$ & $397 \, \mathrm{s}$ \\
 Specific impulse, $I_{sp}$ & $340 \, \mathrm{s}$ & $348 \, \mathrm{s}$ \\
 Thrust, $T$ & $801 \, \mathrm{kN}$ & $934 \, \mathrm{kN}$ \\
 \hline
\end{tabular}
\end{table}

\begin{figure}
  \includegraphics[width=\linewidth]{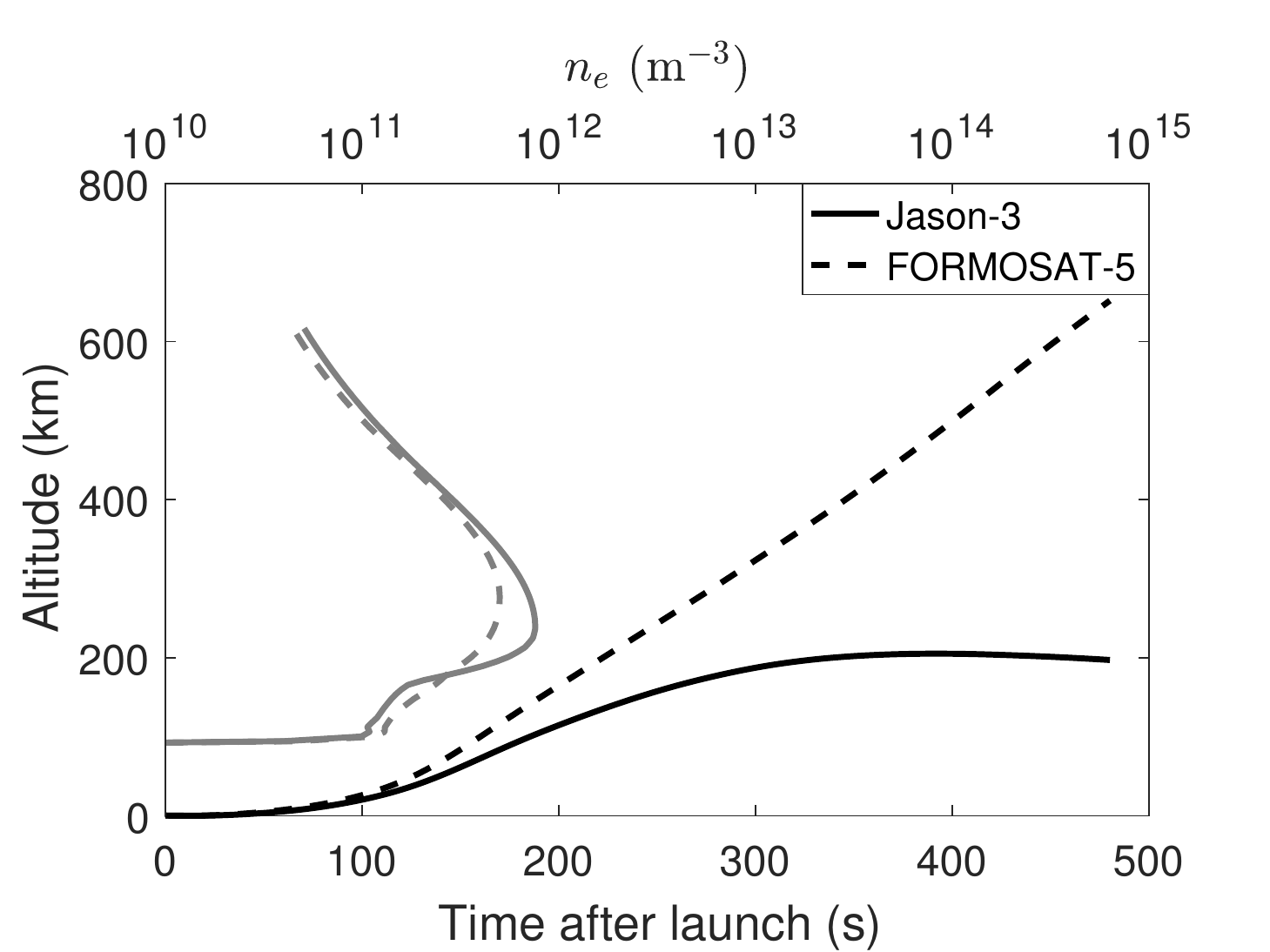}
  \caption{Launch vehicle altitude as a function of time after launch (black lines) for Jason-3 (solid) and FORMOSAT-5 (dashed). The MSIS altitude profiles for electron number density, $n_e$, above the launch site are plotted alongside (grey lines), }
  \label{fig_altitude}
\end{figure}

\begin{figure}
  \includegraphics[width=\linewidth]{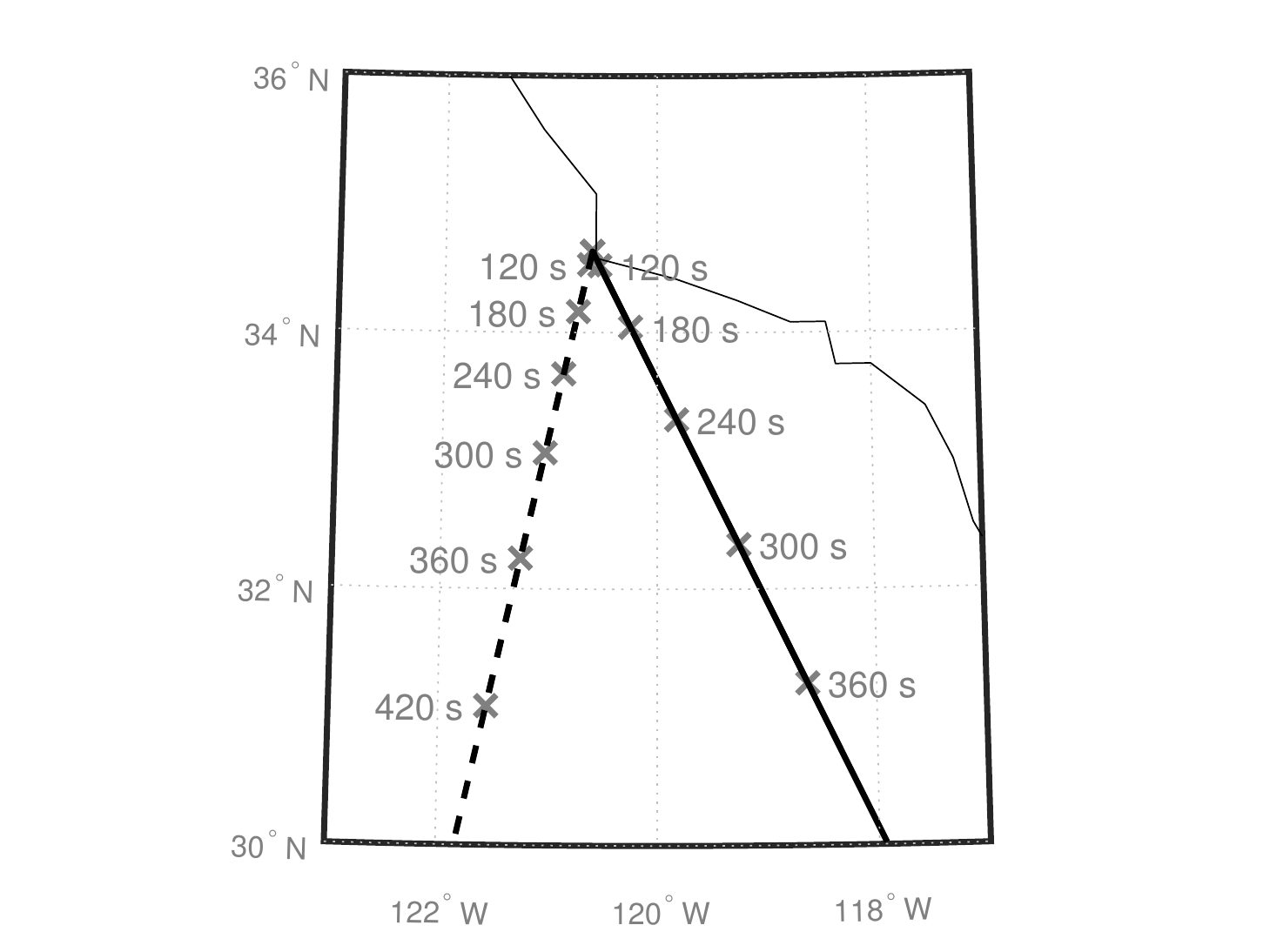}
  \caption{Inferred Jason-3 (solid line) and FORMOSAT-5 (dashed line) ground tracks.}
  \label{fig_trajectory}
\end{figure}

Speed and altitude (Fig.~\ref{fig_altitude}) were determined as functions of time based on launch videos released online by SpaceX, as in Chou et al. \cite{chou2018}. These data were used along with the final orbital inclinations to infer trajectories for both launches, which are plotted in Fig.~\ref{fig_trajectory}. The light FORMOSAT-5 payload enabled the launch vehicle to take an unusually steep trajectory through the F-region of the ionosphere to directly insert the satellite into an orbit at an altitude of $720 \, \mathrm{km}$. By contrast, the more conventional Jason-3 launch inserted the satellite into an initial orbit at $200 \, \mathrm{km}$, remaining below the predicted F-region peak.

\section{Modelling} \label{sec_modelling}

\subsection{GITM}
The Global Ionosphere-Thermosphere Model (GITM) is a general circulation model of the coupled charged and neutral components of the upper atmosphere \cite{ridley2006}. This model uses a finite difference method to solve a set of coupled fluid equations representing a set of ion and neutral species on a 3-dimensional spherical grid which can be stretched in latitude and altitude. Physical and chemical processes in the ionosphere and thermosphere are represented by realistic source terms in the continuity, momentum, and energy equations. Density is explicitly solved for individual major neutral and ion species. Simulations may encompass either the whole globe or a particular region, wherein boundary conditions specify velocities and density and temperature gradients.

A modified version of GITM was developed in this work which incorporates chemical processes, diffusion coefficients, and source terms associated with the rocket exhaust plume. GITM was run in the regional mode to ensure that resolution was sufficient to resolve the features of ionospheric depletions. Boundaries were set at $15^{\circ} \mathrm{N}$, $55^{\circ} \mathrm{N}$, $150^{\circ} \mathrm{W}$, and $90^{\circ} \mathrm{W}$ and an approximate resolution of $0.4^{\circ} \times 0.4^{\circ}$. The model was initialised with input from the empirical International Reference Ionosphere (IRI) and Mass-Spectrometer-Incoherent-Scatter (MSIS) models \cite{bilitza1990,picone2002}.


\subsection{Chemical processes}

In the F-region, \ch{O^+} ions undergo charge exchange reactions with rocket exhaust molecules which proceed ${\sim}2$ to ${\sim}3$ orders of magnitude faster than the corresponding reactions with the \ch{N_2} and \ch{O_2} neutral molecules present in the undisturbed thermosphere \cite{mendillo1981}. The most important reactions are:
\begin{equation}
\mathrm{O^{+} + H_{2}O \to H_{2}O^{+} + O,} \quad  \kappa_1 = 2.42 \times 10^{-15} \, \mathrm{m^3.s^{-1}}, \label{H2O_CE}
\end{equation}
\begin{equation}
\mathrm{O^{+} + H_{2} \to OH^{+} + H,} \quad  \kappa_2 = 1.62 \times 10^{-15} \, \mathrm{m^3.s^{-1}},
\end{equation}
\begin{equation}
\mathrm{O^{+} + CO_{2}}  \begin{matrix}
\mathrm{\overset{0.5}{\rightarrow} O_{2}^{+} + CO} \\ 
\mathrm{\overset{0.5}{\rightarrow} CO_{2}^{+} + O}
\end{matrix} \mathrm{,} \quad  \kappa_3 = 1.08 \times 10^{-15} \, \mathrm{m^3.s^{-1}}, \label{CO2_CE}
\end{equation}
where the reaction rate coefficients $\kappa_1$, $\kappa_2$, and $\kappa_3$ are derived from the measurements summarised by Anicich \cite{anicich2003}, weighted based on the uncertainties reported there. The branching ratios in eq.~\ref{CO2_CE} are chosen based on the results of Lindinger et al. \cite{lindinger1974}, who found that the dominant product ion changed from \ch{O_2^+} to \ch{CO_{2}^{+}} as temperature was increased from $300 \, \mathrm{K}$ to $900 \, \mathrm{K}$. Note that \ch{CO} is taken to be inert with respect to charge exchange with \ch{O^+} \cite{smith1978}. 

The resulting molecular ions undergo dissociative recombination with electrons:
\begin{equation}
\mathrm{H_{2}O^{+} + e^{-}} \begin{array}{ll}
\mathrm{\overset{0.22}{\rightarrow} OH + H } \\
\mathrm{\overset{0.1}{\rightarrow} O + H_2 } \\
\mathrm{\overset{0.68}{\rightarrow} O + H + H } \label{H2O+_DR}
\end{array}
\mathrm{,} \quad  \alpha_1 = 3.0 \times 10^{-13} \, \mathrm{m^3.s^{-1}} ,
\end{equation}
\begin{equation}
\mathrm{OH^{+} + e^{-} \to H + O,} \quad  \alpha_2 = 1.0 \times 10^{-13} \, \mathrm{m^3.s^{-1}} ,
\end{equation}
\begin{equation}
\mathrm{O_{2}^{+} + e^{-} \to O + O,} \quad  \alpha_3 = 2.0 \times 10^{-13} \, \mathrm{m^3.s^{-1}} , \label{O2+_DR}
\end{equation}
where the reaction rate coefficients $\alpha_1$, $\alpha_2$, and $\alpha_3$ are those provided by Mendillo \cite{mendillo1981}. The branching ratios in eq.~\ref{H2O+_DR} are taken from the work of Vejby-Christensen et al. \cite{vejbychristensen1997}.

The above reactions were implemented in the modified version of GITM. Reaction rate coefficients in eq.~\ref{H2O_CE}-\ref{O2+_DR} indicate that the dissociative recombination reactions occur ${\sim}2$ orders of magnitude faster than the charge exchange reactions and the former reactions are therefore treated as instantaneous for the purposes of this study. The enthalpies of reaction for the charge exchange and recombination processes were computed based on the ionisation threshold potentials and enthalpies of formation tabulated by Schunk \& Nagy \cite{schunk2009}. Thermal energy from exothermic reactions was partitioned between reaction products inversely proportional to their masses. Neutral reactants in eq.~\ref{H2O_CE}-\ref{CO2_CE} were implemented as major species. Additionally, those neutral products in eq.~\ref{CO2_CE}-\ref{O2+_DR} which were not previously considered in GITM were implemented as minor species. These changes are summarised in Table~\ref{tab_major_minor_species}.


\begin{table}
\caption{Changes to major and minor neutral species in GITM.}
\label{tab_major_minor_species}
\begin{tabular}{@{}lc}
Major (old) & \ch{O(^3P)}, \ch{O_2}, \ch{N_2}, \ch{N(^4S)}, \ch{NO}, and \ch{He} \\
Minor (old) & \ch{N(^2D)}, \ch{N(^2P)}, \ch{H}, \ch{CO_2}, and \ch{O(^1D)} \\
Major (new) & \ch{O(^3P)}, \ch{O_2}, \ch{N_2}, \ch{N(^4S)}, \ch{NO}, \ch{He}, \ch{H_2O}, \ch{H_2}, and \ch{CO_2} \\
Minor (new) & \ch{N(^2D)}, \ch{N(^2P)}, \ch{H}, \ch{O(^1D)}, \ch{CO}, and \ch{OH} \\
\end{tabular}
\end{table}

\subsection{Diffusion coefficients}
Where they were not previously implemented in GITM, binary gas diffusion coefficients were approximated using the empirical equation and data given by Fuller, Schettler, \& Giddings \cite{fuller1966} in the modified version. According to this equation
\begin{equation}
D_{ab} = \frac{10^{-3} T^{1.75} \left ( \frac{1}{M_a} + \frac{1}{M_b} \right )}{P \left [ \left ( \Sigma V_a \right )^{\frac{1}{3}} + \left ( \Sigma V_b \right )^{\frac{1}{3}} \right ]} ,
\end{equation}
where $D_{ab}$ is the binary diffusion coefficient, $P$ is total pressure (in $\mathrm{atm}$), $M_i$ is molecular weight (in $\mathrm{g.mol}^{-1}$), $T$ is temperature (in $\mathrm{K}$), and $\Sigma V_i$ is diffusion volume (based on tabulated data reflecting structural features of diffusing species $i$). The diffusion coefficient for species $a$ in the gas mixture was estimated using
\begin{equation}
D_{0a} = \frac{1-Y_a}{\sum_{i \ne a} Y_i / D_{ai}} ,
\end{equation}
where $Y_i$ is the molar fraction of species $i$.

\subsection{Rocket exhaust plume source term}



\begin{figure}
  \includegraphics[width=\linewidth]{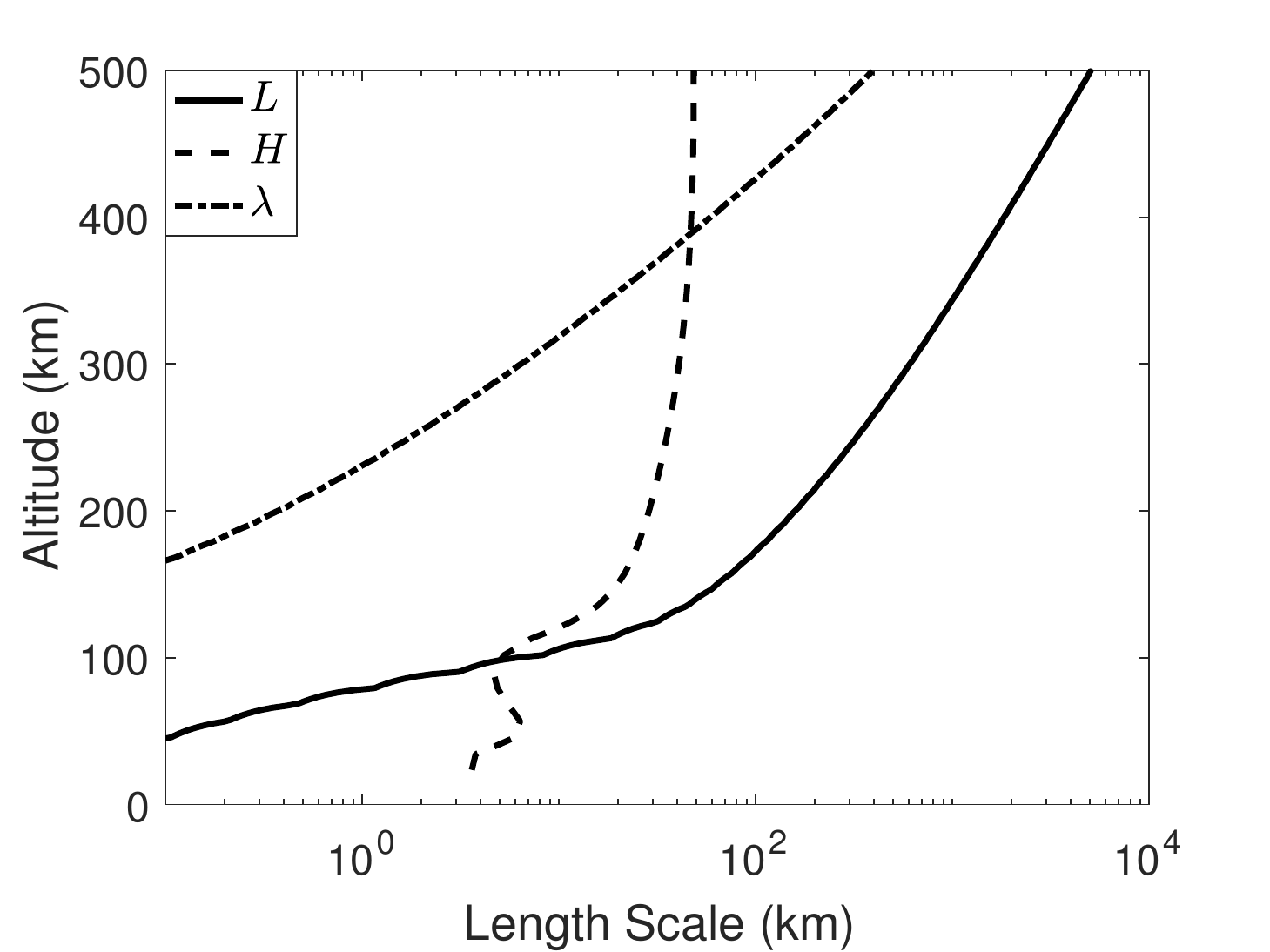}
  \caption{Length scales as functions of altitude for the FORMOSAT-5 launch. Plume length scale, $L$, (solid line), scale height, $H$, (dashed line), and mean free path, (dash-dotted line) are plotted.}
  \label{fig_length_scale}
\end{figure}

Initial diffusion of the rocket exhaust plume in the modified version of GITM was determined analytically using the theory of Bernhardt \cite{bernhardt1979}. This treatment neglected the self-continuum (fluid) and collisionless (free-streaming) initial stages of the exhaust plume expansion \cite{bernhardt1979b}. The plume length scale, $L$, and mean free path, $\lambda$, were assumed to be small compared to the dimensions of the problem. These quantities are plotted for the FORMOSAT-5 launch in Fig.~\ref{fig_length_scale}, based on the hard sphere approximation and gas dynamic theory of rocket plumes developed by Jarvinen et al. \cite{jarvinen1966} respectively. It is noted that the latter theory assumes continuum flow and a homogenous atmosphere, and therefore that $L \ll H$ and $L \ll \lambda$, where $H$ is the scale height of the background atmosphere. Thus, the rocket exhaust species were considered to expand diffusively from rest at the point where they were released in accordance with the approximation given by Bernhardt \cite{bernhardt1979};
\begin{multline}
\sigma_i \left ( x , y , z ; t \right ) = \frac{\dot{N}_{0i}}{\left ( 4 \pi D_{0i} \right )^{\frac{3}{2}}} \exp \left [ -z \left ( \frac{3}{4H} + \frac{1}{2H_i} \right ) \right. \\
\left. -\frac{H^2 \left ( 1 - \exp \left [ - z / \left ( 2 H \right ) \right ] \right )}{D_{0i}t} - \frac{\left ( x^2 + y^2 \right ) \exp \left [ -z/ \left ( 2 H \right ) \right ]}{4D_{0i}t} \right. \\
\left. -\left ( \frac{1}{H} -\frac{1}{H_i} \right )^2 \frac{D_{0i} t \exp \left [ z / \left ( 2H \right ) \right ]}{4} \right ] , \label{gas_diffusion_point}
\end{multline}
where $\sigma_i$ is the source term, $\dot{N}_{0i}$ is the molecular flow rate, and $D_{0i}$ is the diffusion coefficient for species $i$. Here we define $H = k T / \left ( m g \right )$ as the scale height of the background atmosphere and $H_i = k T / \left ( m_i g \right )$ as the scale height of species $i$, where $k$ is the Boltzmann constant, $T$ is the temperature, $m$ is the mean molecular mass of the background atmosphere, $m_i$ is the molecular mass of species $i$, and $g$ is gravitational acceleration. In the implementation of eq.~\ref{gas_diffusion_point} in GITM, the $x$, $y$, and $z$ coordinates are approximated by the zonal, meridional, and vertical displacements with respect to the point of release respectively. Thus, $x \approx r_0 \cos \theta_0 \left ( \phi - \phi_0 \right )$, $y \approx r_0 \left ( \theta - \theta_0 \right )$, and $z = r -  r_0$, where $\theta_0$, $\phi_0$, and $r_0$ are the spherical coordinates of the point of release. The diffusion time, $t$, represents the difference between the time between the the exhaust being released and the corresponding addition of the source term in the numerical model. A sufficiently large value of $t$ was required to avoid failure of the GITM numerical solver, as excessive gradients in rocket exhaust species led to the occurence of negative densities. Chemical reactions during that time period were neglected in this treatment.

Estimates of $\dot{N}_{0i}$ for each species were obtained based on the manufacturer data stated in Table~\ref{tab_F9_data}. Simulations were run for two different methods of determining the chemical composition of the exhaust; firstly assuming complete combustion and secondly using Chemical Equilibrium with Applications (CEA) software \cite{gordon1996}. In the latter case a combustion pressure of $6.895 \, \mathrm{MPa}$, combustion temperature of $3670 \, \mathrm{K}$, nozzle expansion ratio of $165$, and oxidiser to fuel ratio of $2.56$ were chosen. Conditions at the throat and exit were considered, representing the bounding cases for combustion. The resulting molecular flow rates are summarised in Table~\ref{tab_mfr_data}. It was assumed that the heating of neutral gasses in the thermosphere as the exhaust gasses as they came to rest was equal to the kinetic energy of the latter as they exited the rocket (in a reference frame stationary with respect to Earth), such that
\begin{equation}
\dot{Q} = \dot{N}_{0i} \left ( v - I_{sp} g_0 \right )^2 ,
\end{equation}
where $\dot{Q}$ is the heating rate, $v$ is the speed of the rocket, $I_{sp}$ is the specific impulse of the rocket, and $g_0$ is the reference gravitational acceleration. The source was added until $480 \, \mathrm{s}$ since launch. This preceded the actual termination of the second stage burn by approximately $60 \, \mathrm{s}$ in each case, though in the FORMOSAT-5 launch case the rocket was above the top of the simulation domain by time.

\begin{table*}
\begin{minipage}{130mm}
\caption{Molecular flow rates, $\dot{N}_{0i}$, for the Falcon 9 second stage.}
\label{tab_mfr_data}
\begin{tabular}{@{}lcccc}
 \hline
 & \ch{H_2O} & \ch{H_2} & \ch{CO_2} & \ch{CO} \\
 \hline
 Complete combustion, v1.1 & $2.47 \times 10^{27} \, \mathrm{s}^{-1}$ & $0 \, \mathrm{s}^{-1}$ & $2.28 \times 10^{27} \, \mathrm{s}^{-1}$ & $0 \, \mathrm{s}^{-1}$ \\
 Complete combustion, v1.2 & $2.81 \times 10^{27} \, \mathrm{s}^{-1}$ & $0 \, \mathrm{s}^{-1}$ & $2.59 \times 10^{27} \, \mathrm{s}^{-1}$ & $0 \, \mathrm{s}^{-1}$ \\
 CEA throat, v1.2 &  $2.09 \times 10^{27} \, \mathrm{s}^{-1}$ & $5.07 \times 10^{26} \, \mathrm{s}^{-1}$ & $9.69 \times 10^{26} \, \mathrm{s}^{-1}$ & $1.94 \times 10^{27} \, \mathrm{s}^{-1}$ \\
 CEA exit, v1.2 &  $2.21 \times 10^{27} \, \mathrm{s}^{-1}$ & $1.02 \times 10^{27} \, \mathrm{s}^{-1}$ & $1.89 \times 10^{27} \, \mathrm{s}^{-1}$ & $1.43 \times 10^{27} \, \mathrm{s}^{-1}$ \\
 \hline
\end{tabular}
\end{minipage}
\end{table*}

\section{Results} \label{sec_results}

\begin{figure*}
\centering
\subfloat[Jason-3]{
\includegraphics[width=\linewidth]{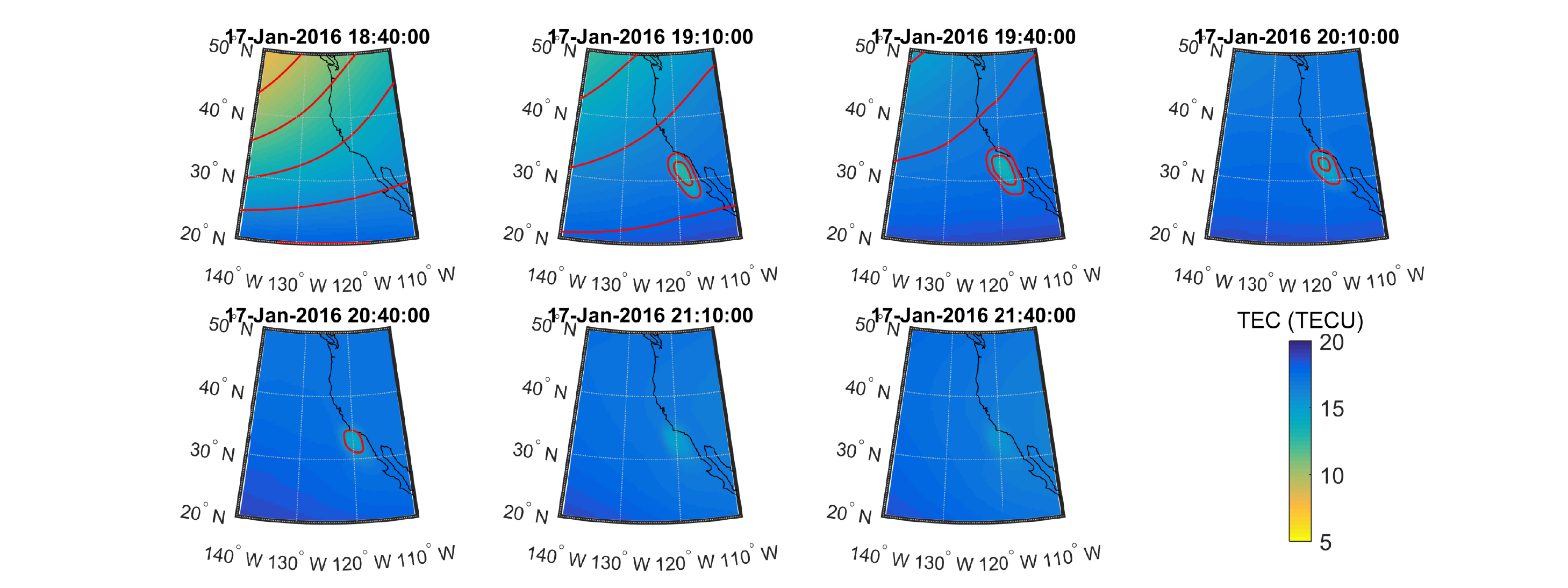}
}
\hspace{0mm}
\subfloat[FORMOSAT-5]{
\includegraphics[width=\linewidth]{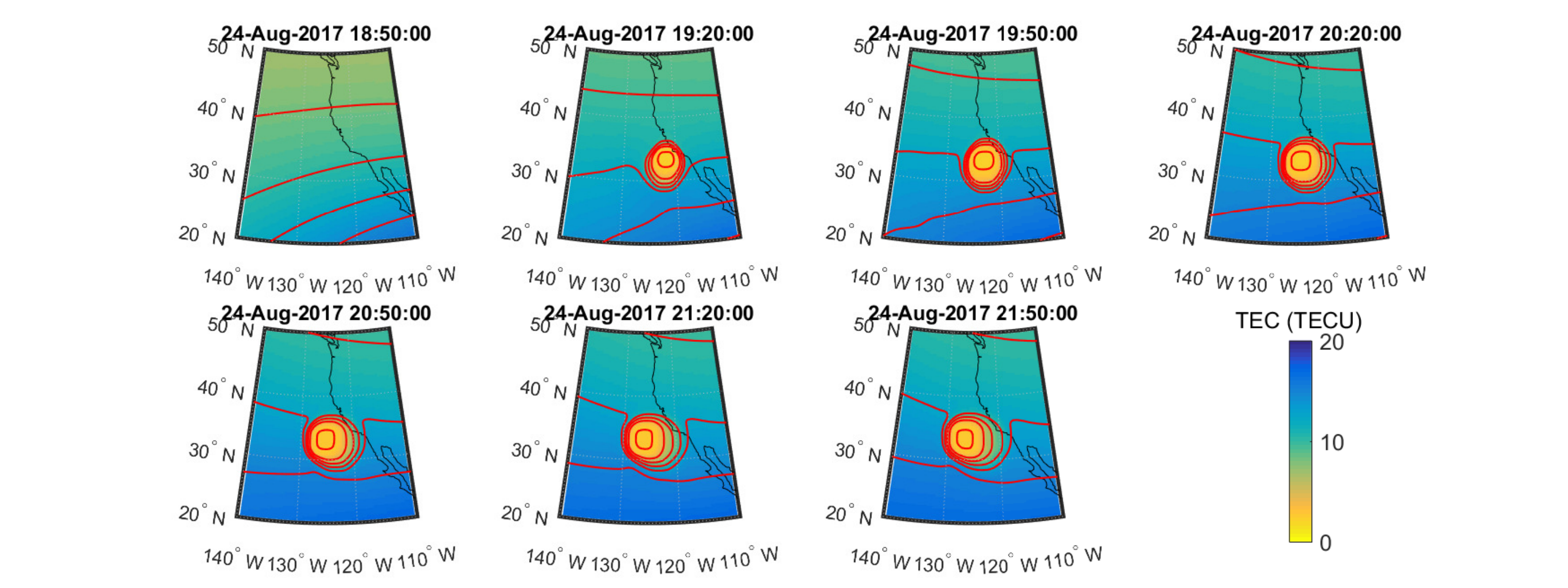}
}
\caption{TEC maps from GITM simulation of the ionospheric depletion following rocket launches assuming complete combustion.}
\label{fig_GITM_TEC_map_data}
\end{figure*}

Ionospheric depletions due to the Jason-3 and FORMOSAT-5 launches were modelled using GITM assuming complete combustion of the propellant. Evolution of the depletions over the $3$~hours following launch in each simulation are shown in the TEC maps in Fig.~\ref{fig_GITM_TEC_map_data}. The two launches coincided with periods of relatively low geomagnetic activity ($\mathrm{Kp} = 1-$ and $0+$ respectively) and solar activity ($\mathrm{F}10.7 = 101 \, \mathrm{ SFU}$ and $79 \, \mathrm{ SFU}$ respectively). Nevertheless, substantially higher background TEC values were computed for the Jason-3 launch ($15.5 \, \mathrm{ TECU}$ at Vandenberg Air Force Base) than for the FORMOSAT-5 launch ($11.8 \, \mathrm{ TECU}$).



\begin{figure}
\includegraphics[width=\linewidth]{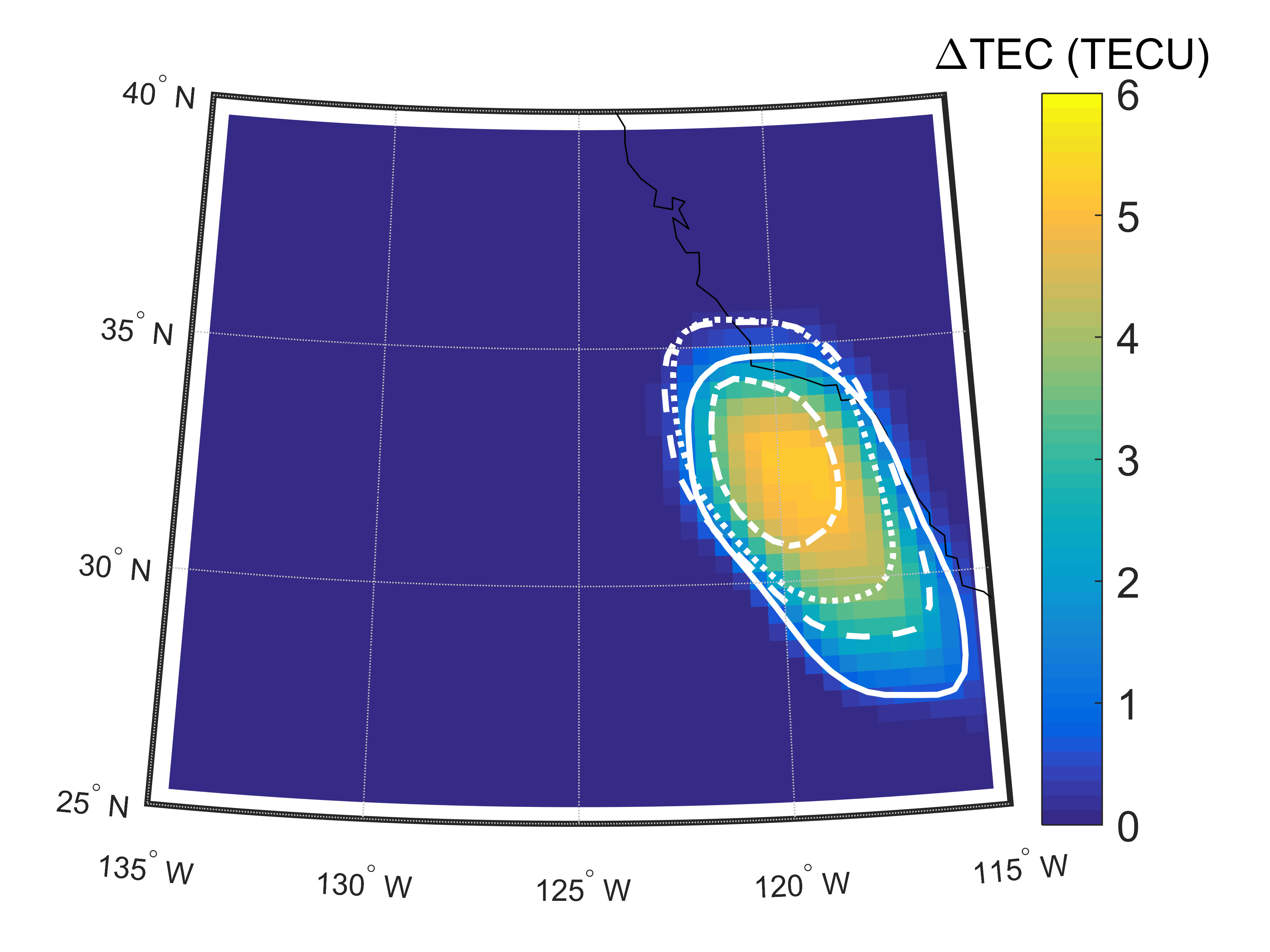}
\caption{TEC depletion map for 19:50 UT 24 August 2017, the time when the maximum TEC depletion was recorded, from GITM simulations of the launch of Jason-3 assuming complete combustion. Contours of $\Delta \mathrm{TEC} = 1 \mathrm{ TECU}$ are also plotted for 19:40 UT (solid line), 20:40 UT (dashed line), 21:40 UT (dotted line), and 22:40 UT (dash-dotted line).}
\label{fig_GPS_TEC_diff_J3_data}
\end{figure}

\begin{figure}
\includegraphics[width=\linewidth]{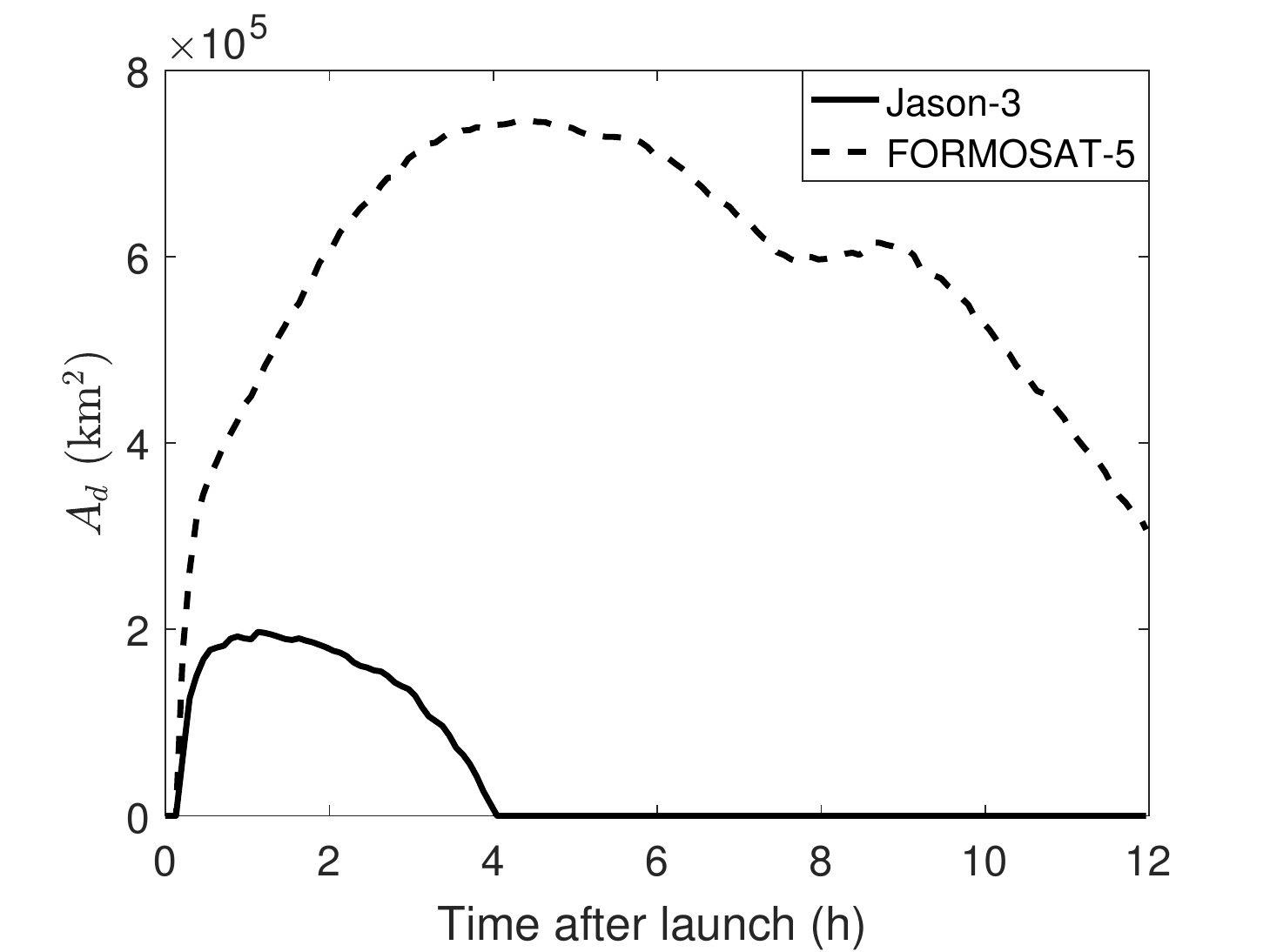}
\caption{Evolution of the horizontal spatial extent of the ionosphere depletion, $A_d$, in simulations of the Jason-3 (dashed line) and FORMOSAT-5 (dash-dotted line) launches.}
\label{fig_hole_area}
\end{figure}

In the Jason-3 launch case, the simulation indicated that an elongated ionospheric depletion formed rapidly, aligned with the launch vehicle trajectory.  The difference in TEC between a simulation in which no rocket exhaust gases were added to the atmosphere and the simulation including these gases was taken to represent the ionospheric depletion and is shown in Fig.~\ref{fig_GPS_TEC_diff_J3_data}. The horizontal spatial extent of the depletion is represented by $A_d$, the area of the Earth's surface over which this difference exceeds $1 \, \mathrm{ TECU}$, which is plotted as a function of time in Fig.~\ref{fig_hole_area}. A maximum depletion of $5.26 \, \mathrm{ TECU}$ was recorded at 19:50 UT when $A_d = 1.97 \times 10^5 \, \mathrm{km}^2$. Over subsequent hours, the magnitude and spatial extent of the depletion decreased and it became more circular. Its centre moved northwestward, towards the launch site. The depletion in TEC was greater on the side of the depletion nearer to the launch site, as the slower speed of the launch vehicle earlier in its trajectory produced greater concentrations of exhaust gases while flight was approximately level from $300 \, \mathrm{s}$ after launch onwards.

\begin{figure}
\includegraphics[width=\linewidth]{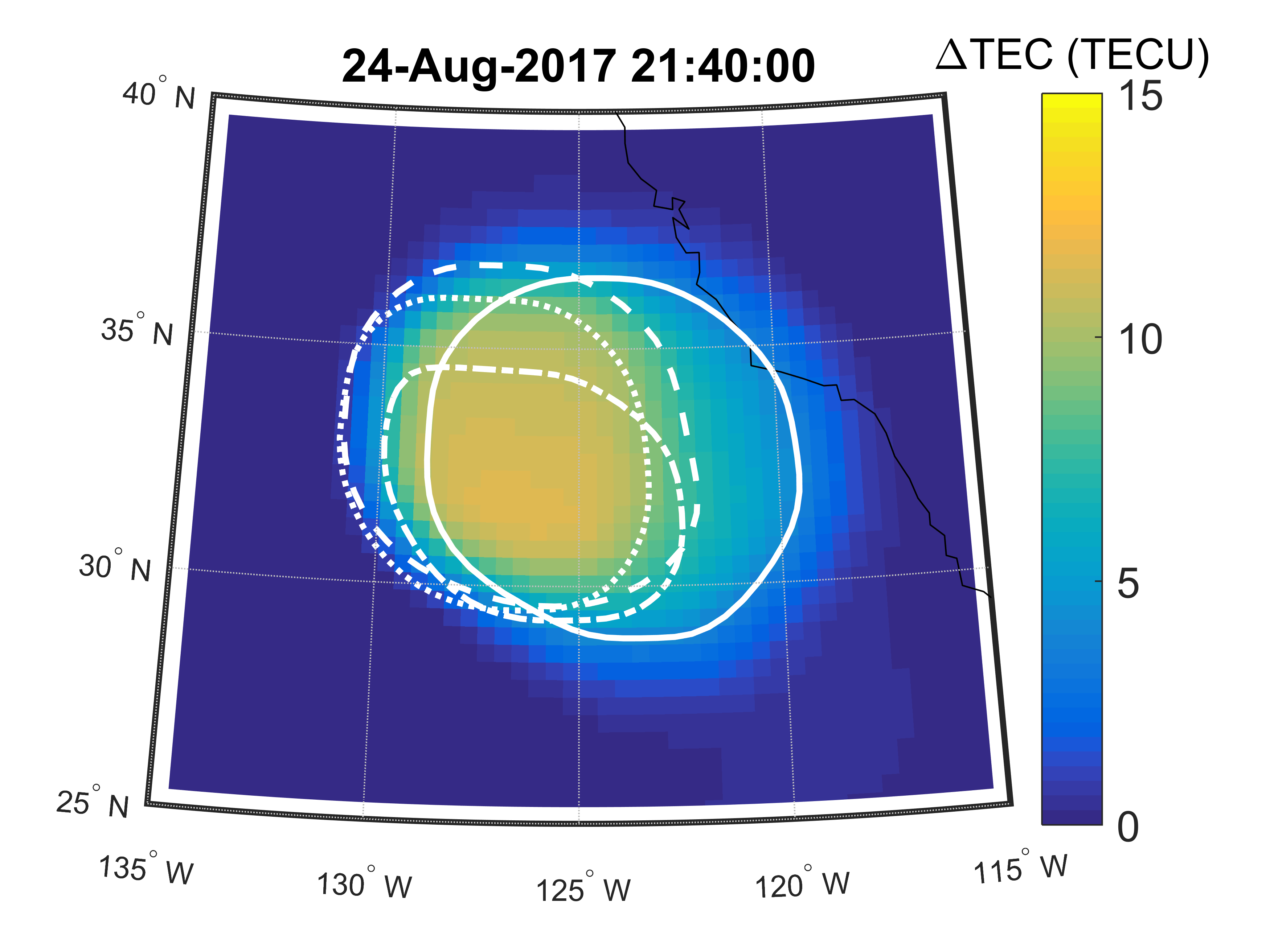}
\caption{TEC depletion map for 21:40 UT 24 August 2017, the time when the maximum TEC depletion was recorded, from GITM simulations of the launch of FORMOSAT-5 assuming complete combustion. Contours of $\Delta \mathrm{TEC} = 5 \mathrm{ TECU}$ are also plotted for 20:50 UT (solid line), 22:50 UT (dashed line), 00:50 UT (dotted line), and 02:50 UT (dash-dotted line).}
\label{fig_GPS_TEC_diff_data}
\end{figure}

By contrast, in the FORMOSAT-5 launch case, a large, persistent, and nearly circular ionospheric depletion formed. The decrease in TEC in this case relative to that without addition of rocket exhaust gases is shown in Fig.~\ref{fig_GPS_TEC_diff_data}, with a maximum depletion of $11.5 \, \mathrm{ TECU}$ which was recorded at 21:40 UT. However, the maximum spatial extent occurred after further diffusion at 23:20 UT, when $A_d = 7.46 \times 10^5 \, \mathrm{km}^2$. Despite the release of similar quantities of rocket exhaust as in the previous case, much larger depletions occurred across a much larger geographical region in this instance. The depletion initially moved westward for several hours before coming to a halt and beginning to move back eastward. This movement was attributed to transport of the exhaust by neutral winds. A trail of reduced TEC resulted as the ionosphere, the upper regions of which are largely constrained to move along magnetic field lines, recovered. The ionospheric depletion continued to exist following sunset (which occurred at 02:38 UT August 25 at Vandenberg Air Force Base), though absolute depletion values decreased at night.

\begin{figure}
\includegraphics[width=\linewidth]{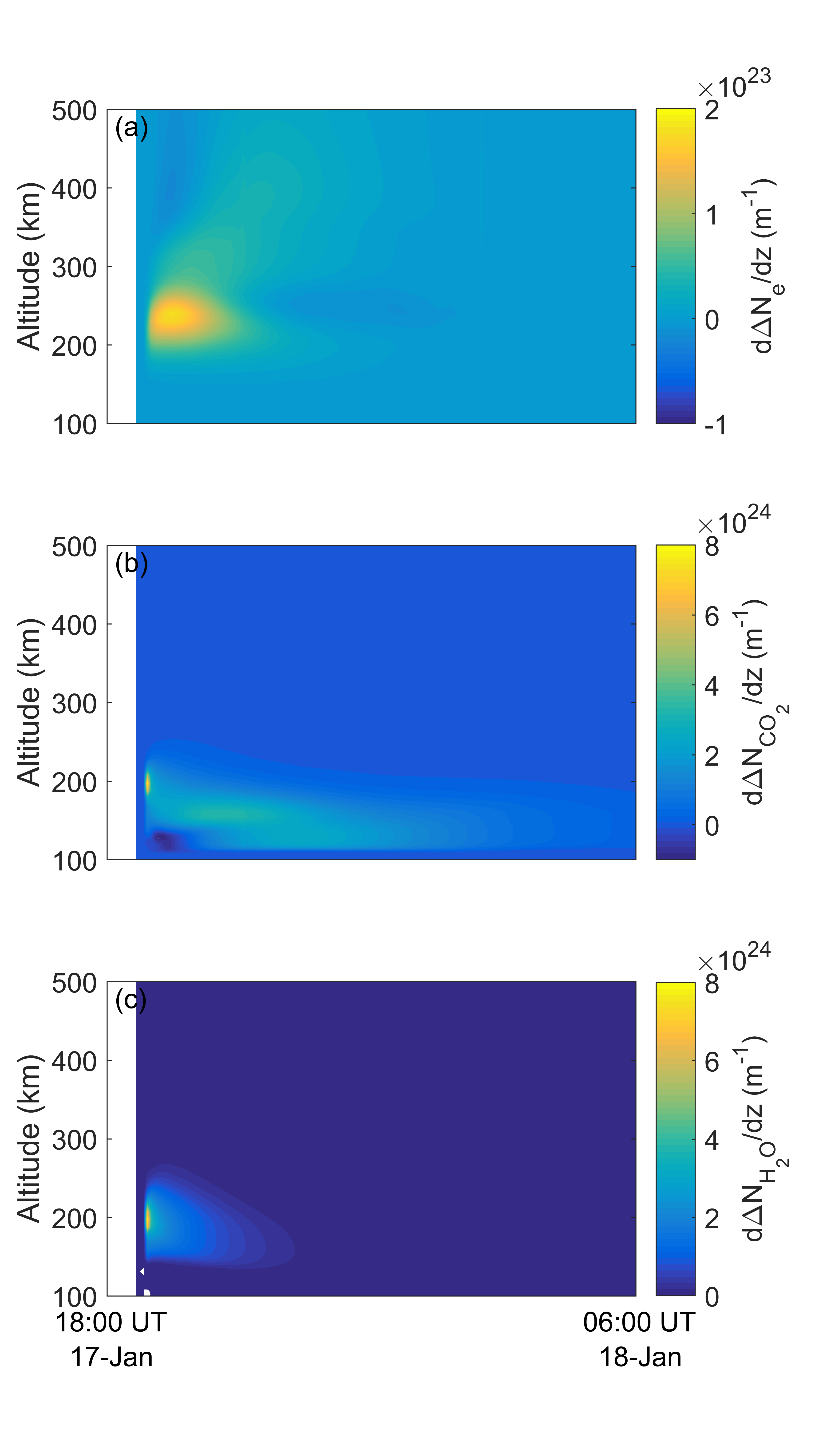}
\caption{Distribution of differences in (a) $N_e$, (b) $N_{\mathrm{CO_2}}$, and (c) $N_{\mathrm{H_2O}}$ as a function of altitude and time due to the introduction of rocket exhaust in the GITM simulation of the Jason-3 launch.}
\label{fig_J3_altitude_profiles}
\end{figure}

\begin{figure}
\includegraphics[width=\linewidth]{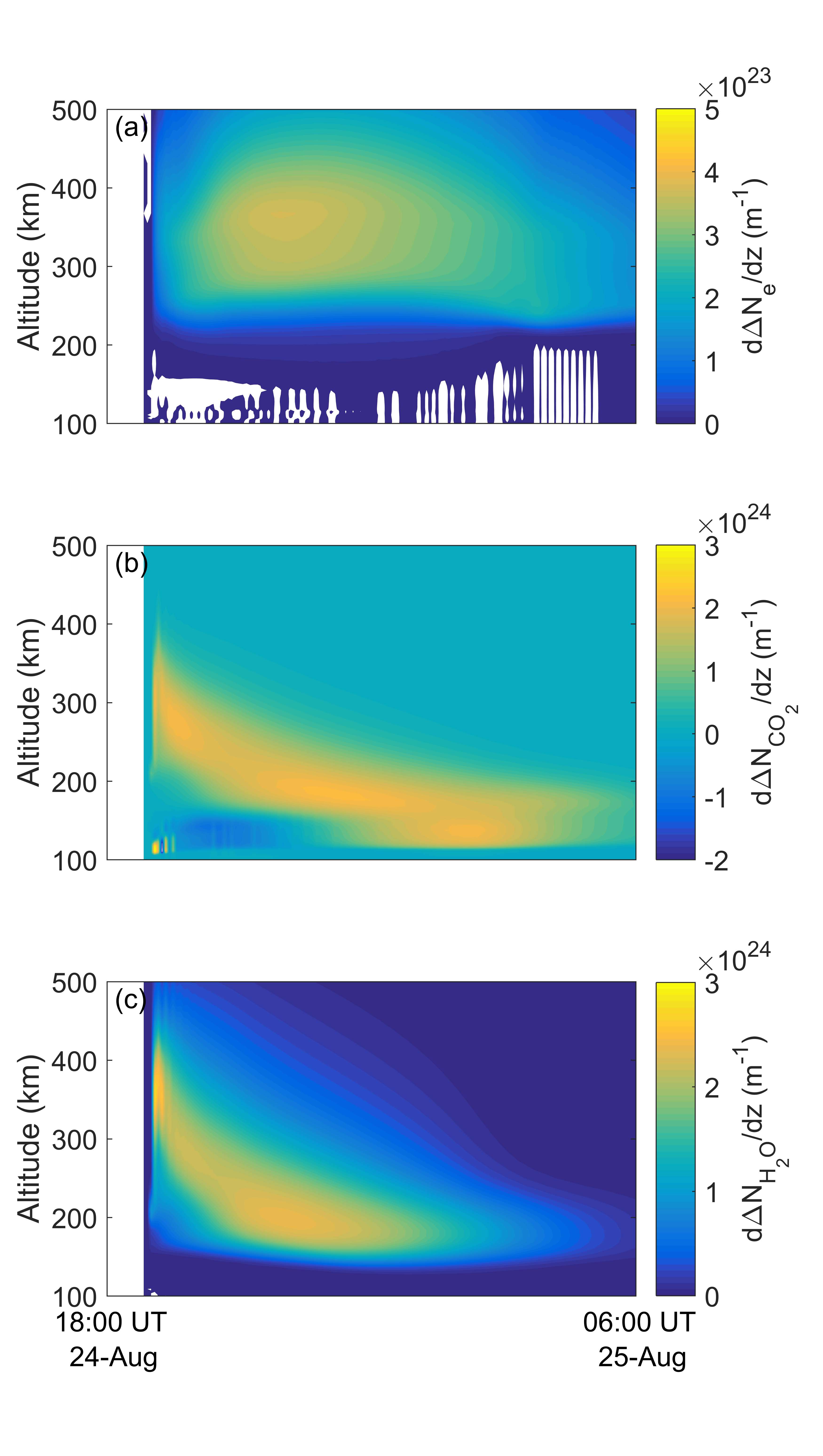}
\caption{Distribution of differences in (a) $N_e$, (b) $N_{\mathrm{CO_2}}$, and (c) $N_{\mathrm{H_2O}}$ as a function of altitude and time due to the introduction of rocket exhaust in the GITM simulation of the FORMOSAT-5 launch.}
\label{fig_F5_altitude_profiles}
\end{figure}

The altitude distributions of changes in the ionosphere and thermosphere due to the rocket launches provide insight into the difference between the two cases above. These distributions are plotted as functions of time for the Jason-3 and FORMOSAT-5 launch cases in Fig.~\ref{fig_J3_altitude_profiles} and Fig.~\ref{fig_F5_altitude_profiles} respectively. We adopt the convention that positive $\Delta N_e$ represents a decrease in the number of electrons in the simulation domain due to the introduction of rocket exhaust, whereas positive $\Delta N_i$ represents an increase in the number of any other species $i$. In the Jason-3 case, the increases in \ch{CO_2} and \ch{H_2O} number were initially concentrated around $200 \, \mathrm{km}$ altitude. The centres of their non-hydrostatic altitude distributions fell and they broadened through diffusion. Overall, increases in \ch{CO_2} and \ch{H_2O} due to rocket exhaust decreased with time as a result of the charge exchange and dissociative recombination reactions in eq.~\ref{H2O_CE}-\ref{O2+_DR}. In the FORMOSAT-5 case, the initial increases in \ch{CO_2} and \ch{H_2O} number occurred across a broader range of altitudes. The centres of their altitude distributions fell over the course of several hours, with losses occurring at the lower boundary of the simulation. In both cases artificial changes in the distributions were found near the lower boundary.

The altitude distributions of increases in \ch{CO_2} and \ch{H_2O} number due to the rocket exhaust discussed above help explain the distribution of the reduction in the electron number. In the Jason-3 case the decrease in electron number was centred around $230 \, \mathrm{km}$ altitude, attaining its greatest value about $1$~hour after launch. These observations reflected the vertical diffusion of rocket exhaust upward into the \ch{O^+} dominated F-region with which it could react. Thereafter, the fall of the rocket exhaust gases and their consumption through charge exchange and dissociative recombination reactions resulted in a reduction in the electron loss rate, allowing the ionosphere to recover. The decrease in electron number in the FORMOSAT-5 case was distributed over a broad range of altitudes reflecting the distribution of increases in \ch{CO_2} and \ch{H_2O}. Even after the exhaust gasses had fallen to lower altitudes, the depletion remained at higher altitudes due to the combination of low electron production rates and the importance of diffusion from lower altitudes.

\begin{figure*}
\centering
\subfloat[Jason-3 complete combustion]{
\includegraphics[scale=.5]{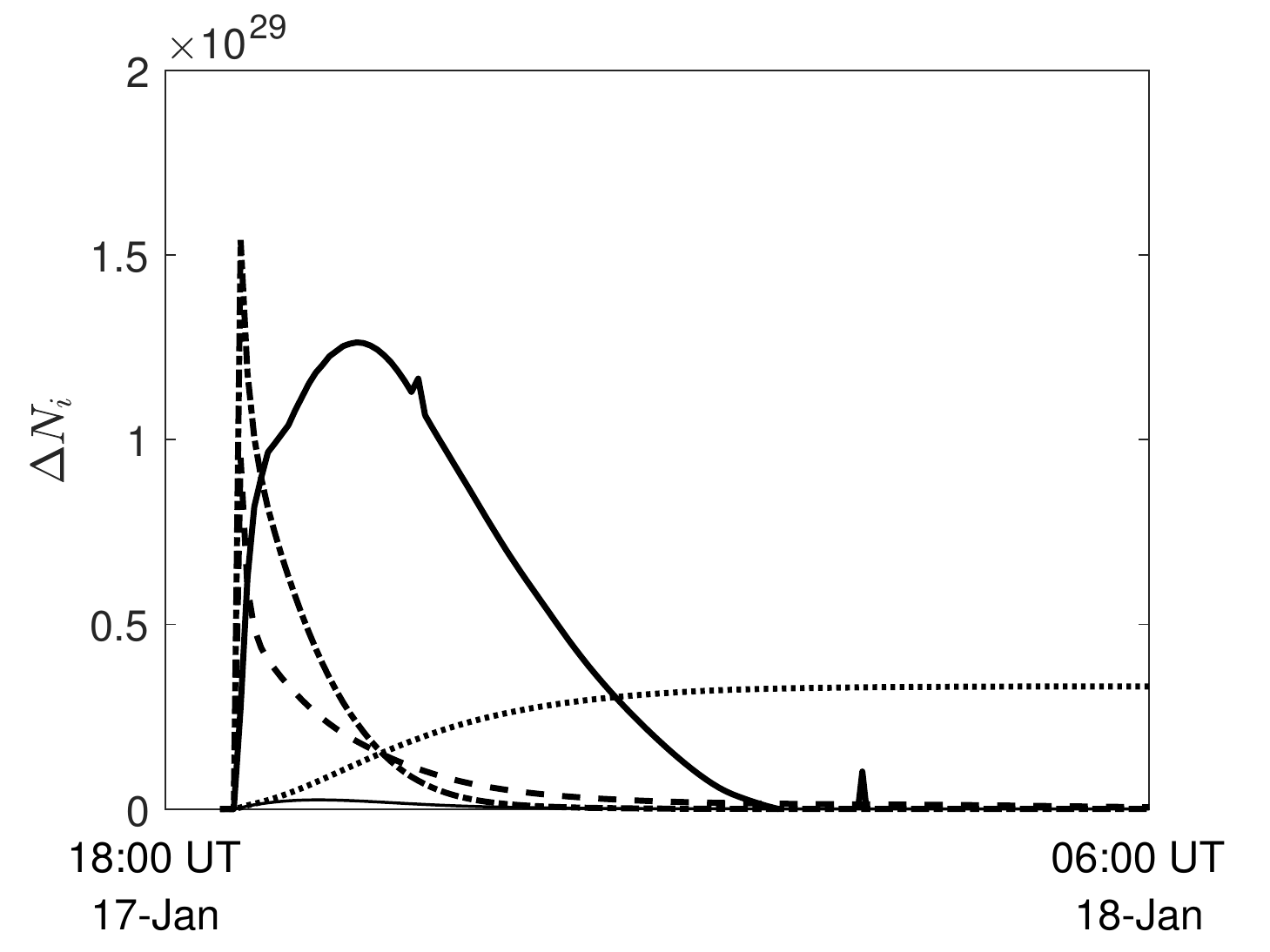}
\label{subfig_J3_CC_Number_data}
}
\hspace{0mm}
\subfloat[FORMOSAT-5 complete combustion]{
\includegraphics[scale=.5]{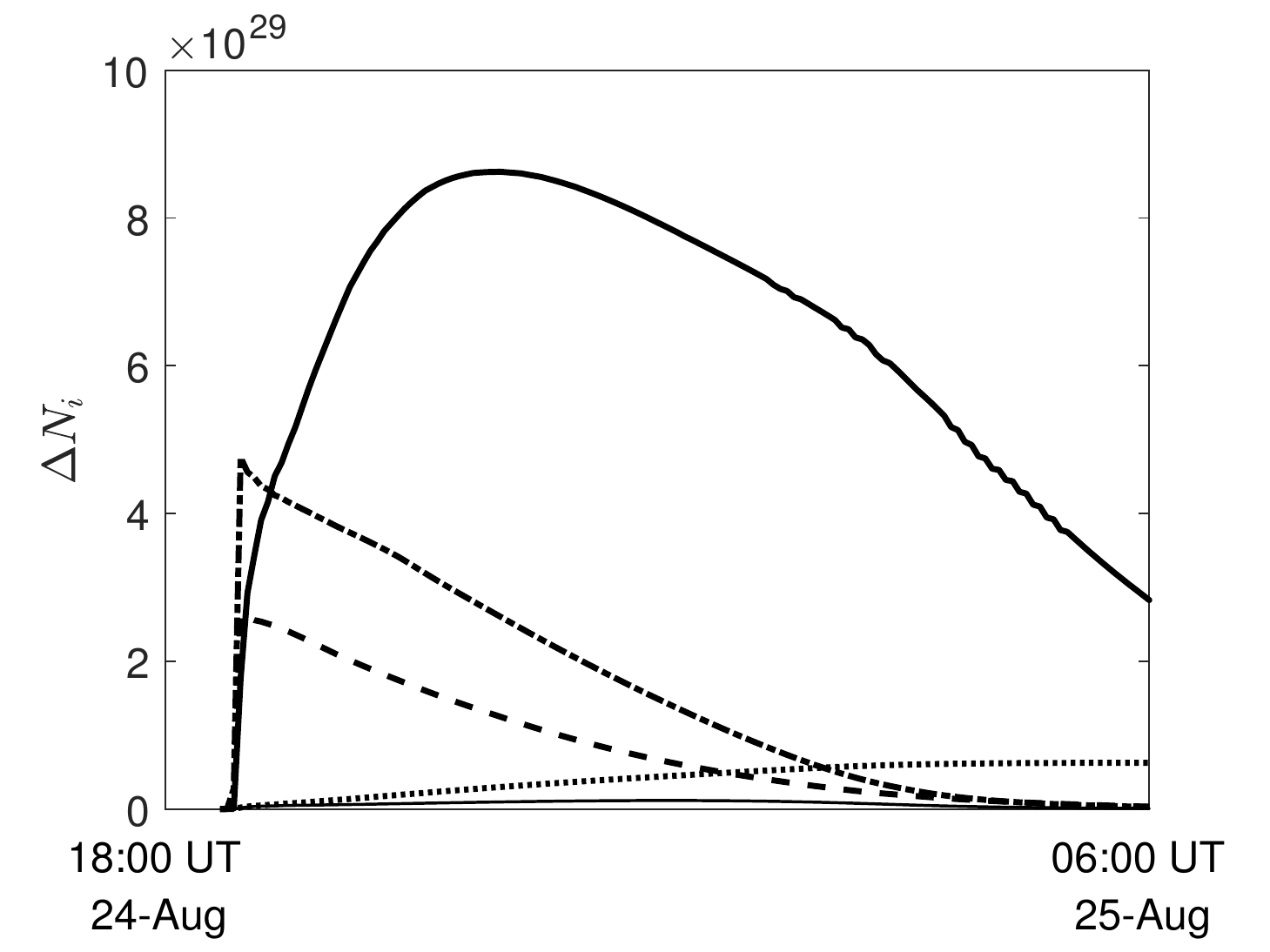}
\label{subfig_F5_CC_Number_data}
}
\hspace{0mm}
\subfloat[FORMOSAT-5 CEA throat]{
\includegraphics[scale=.5]{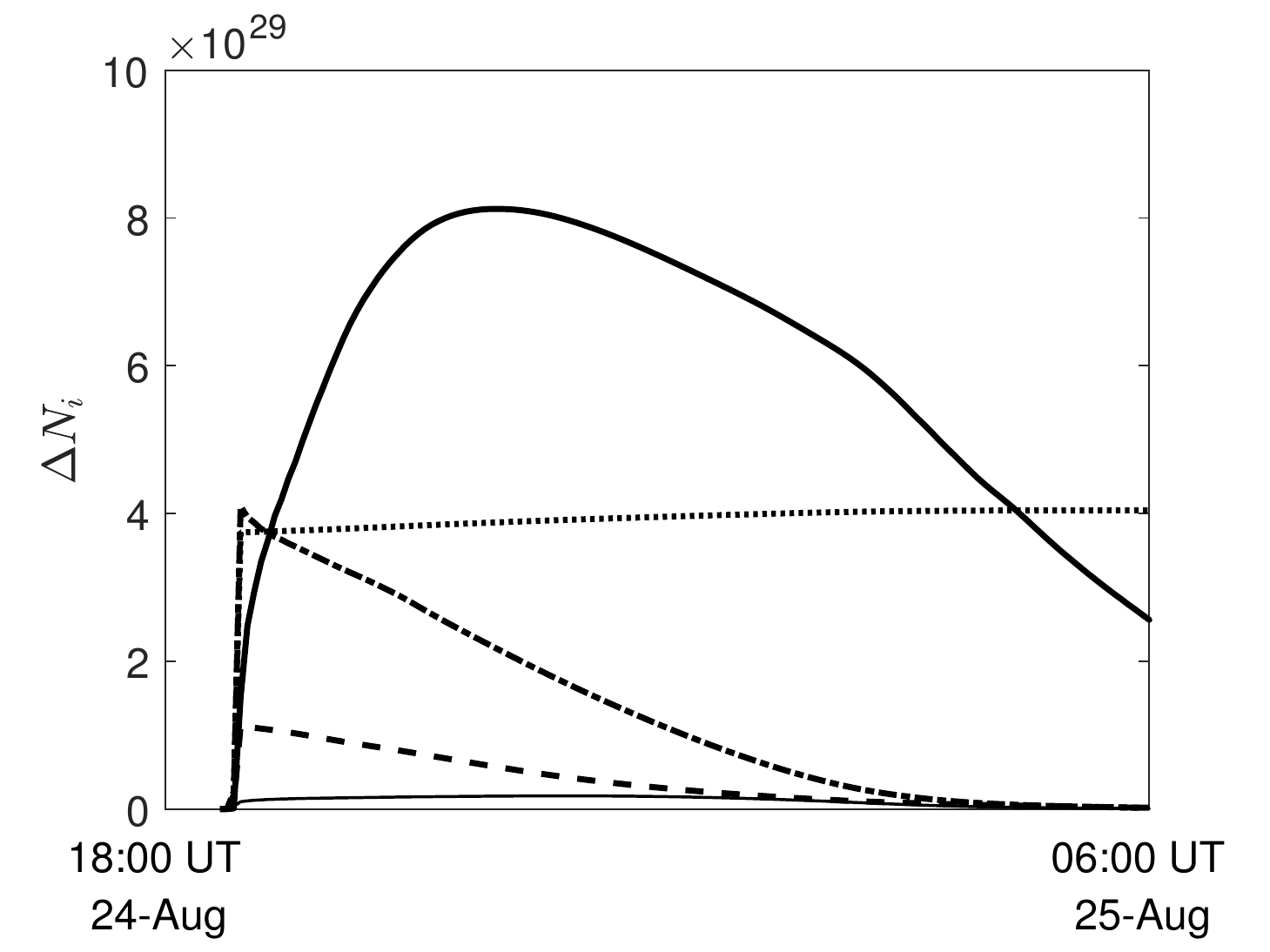}
\label{subfig_F5_throat_Number_data}
}
\hspace{0mm}
\subfloat[FORMOSAT-5 CEA exit]{
\includegraphics[scale=.5]{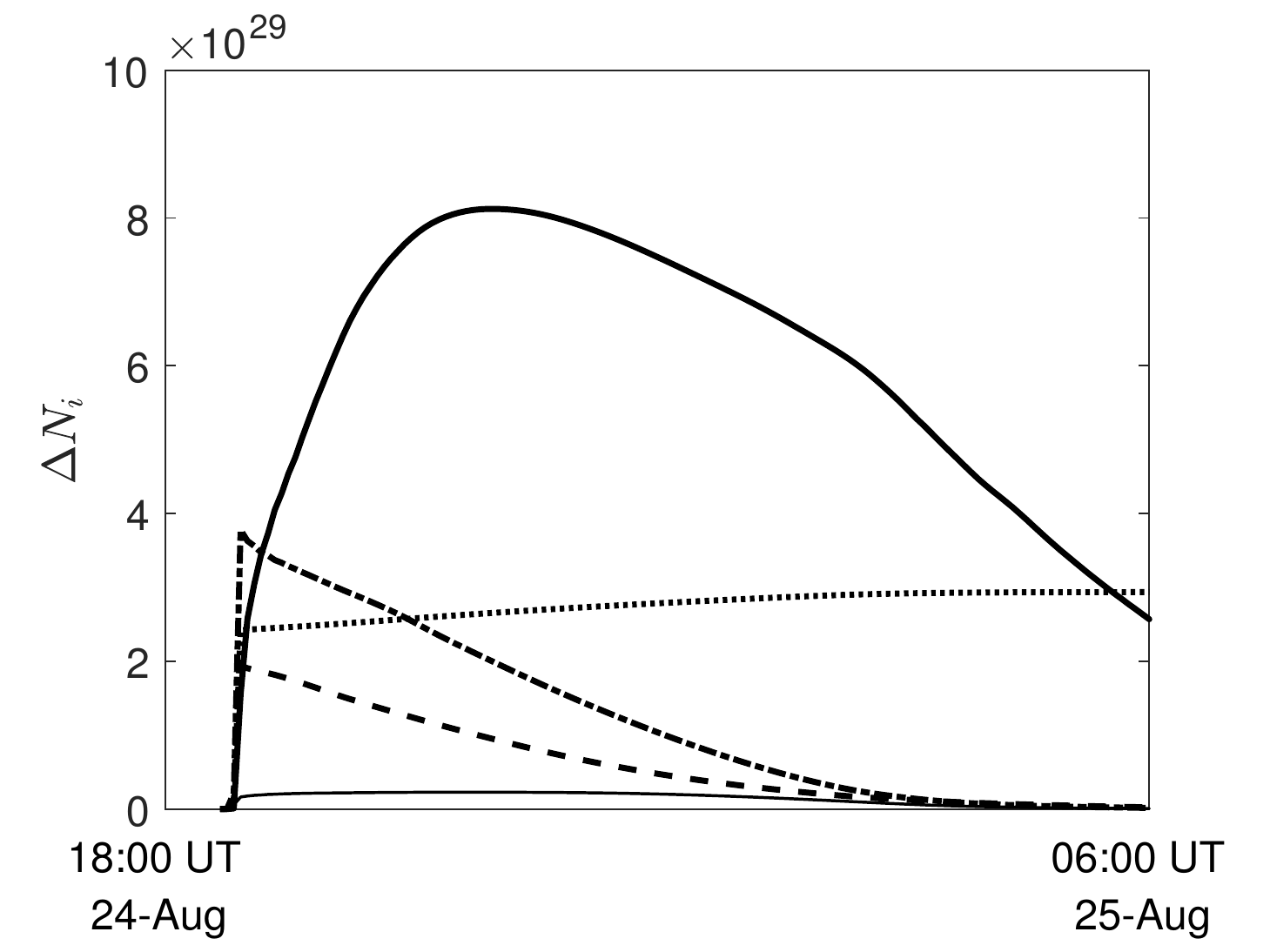}
\label{subfig_F5_exit_Number_data}
}
\caption{Differences in $N_e$ (solid line, multiplied by a factor of $10$), $N_{\mathrm{CO_2}}$ (dashed line), $N_{\mathrm{CO}}$ (dotted line), $N_{\mathrm{H_2O}}$ (dash-dotted line), and $N_{\mathrm{H_2}}$ (thin line)} above $200 \, \mathrm{ km}$ altitude as a function of time due to the introduction of rocket exhaust.
\label{fig_Number_data}
\end{figure*}

Total changes in the number of electrons and each of the rocket exhaust species considered in the simulation are plotted in Fig.~\ref{fig_Number_data}. As seen in Fig.~\ref{fig_J3_altitude_profiles} and Fig.~\ref{fig_F5_altitude_profiles}, reductions in electron number density are largely confined to altitudes greater than $200 \, \mathrm{km}$, therefore only particles above this altitude are considered. Differences in \ch{CO_2} and \ch{H_2O} declined rapidly in the Jason-3 launch case, shown in Fig~\ref{subfig_J3_CC_Number_data}, and more slowly in the Formosat-5 launch case, shown in Fig~\ref{subfig_F5_CC_Number_data}, reflecting the combined effects of diffusion and gravity. Charge exchange and dissociative recombination reactions led to small differences in \ch{CO} and \ch{H_2}, even though these were not present in the exhaust. The total number of electrons lost in this region due to the rocket exhaust was much higher in the FORMOSAT-5 launch case than the Jason-3 launch case (a factor of $\approx 6.8$ greater maximum $\Delta N_e$).

Simulations of the FORMOSAT-5 launch case were also conducted using the rocket exhaust composition computed by CEA at the throat and at the exit. The resulting changes in the numbers of different species are shown in Fig.~\ref{subfig_F5_throat_Number_data} and Fig~\ref{subfig_F5_exit_Number_data}. In each case the number of \ch{CO_2} and \ch{H_2O} molecules added by the rocket were significantly reduced, while the number of \ch{CO} and \ch{H_2} molecules (previously only present as products of reactions between the exhaust gasses and ionosphere) greatly increased. However, it was found that the overall decrease in the number of electrons was not sensitive to these changes in exhaust composition, with less than $6\%$ change in its greatest value.

\section{Observational data} \label{sec_observation}

\begin{figure*}
\centering
\subfloat[Jason-3]{
\includegraphics[width=\linewidth]{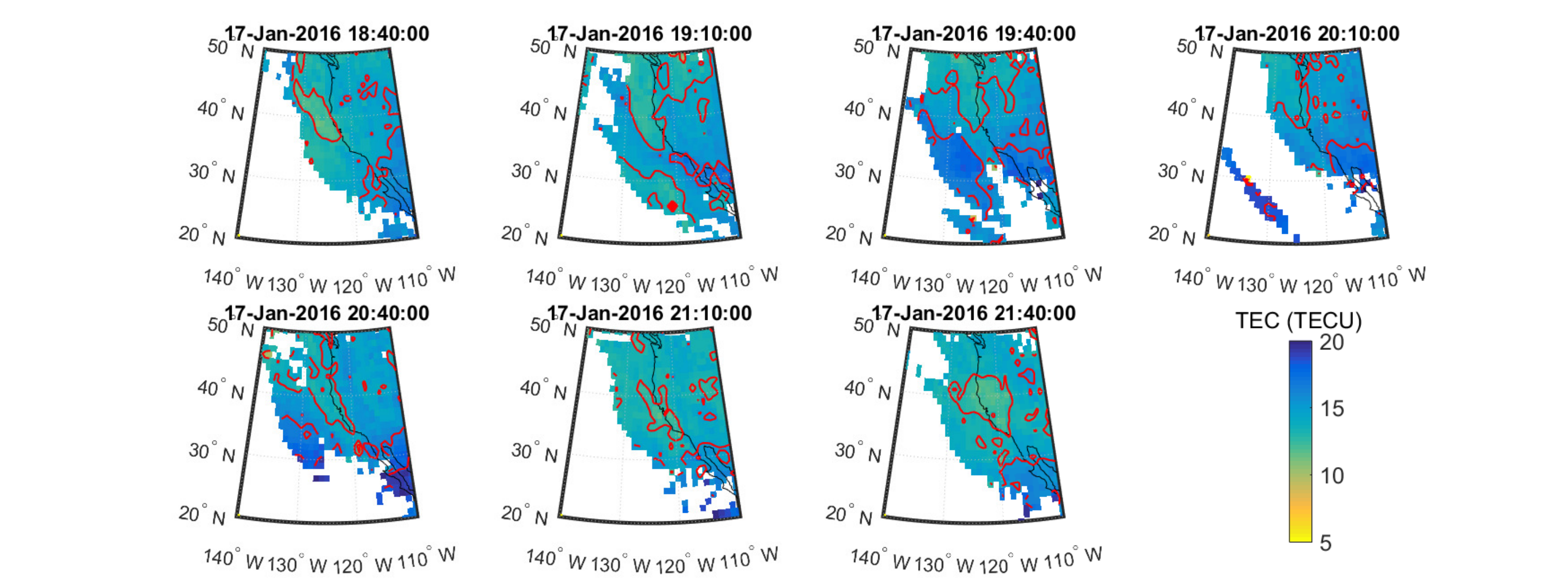}
\label{subfig_MAPGPS_J3_TEC_data}
}
\hspace{0mm}
\subfloat[FORMOSAT-5]{
\includegraphics[width=\linewidth]{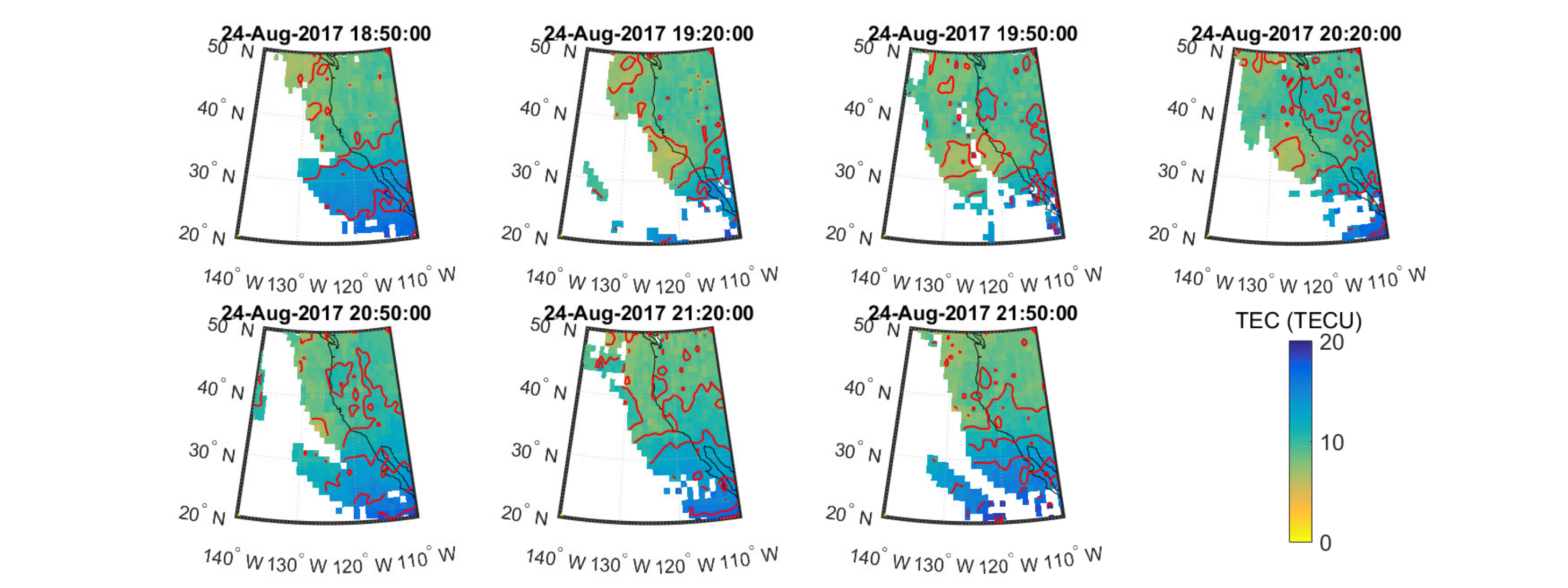}
\label{subfig_MAPGPS_F5_TEC_data}
}
\caption{TEC maps derived from ground-based GNSS observations showing the ionospheric depletion following rocket launches.}
\label{fig_MAPGPS_TEC_data}
\end{figure*}

TEC data maps based on GNSS observations following the FORMOSAT-5 launch show ionospheric depletions similar to those seen in the GITM simulation. Maps with $1^{\circ} \times 1^{\circ}$ resolution, which were derived from MAPGPS software \cite{rideout2006}, are plotted in Fig.~\ref{fig_MAPGPS_TEC_data}. While such measurements contain D-region and plasmasphere contributions to TEC that are excluded from the GITM simulation, these are typically small during the daytime \cite{yizengaw2008}. A region of reduced TEC was observed to form near the launch site and subsequently move westward, as seen in the simulation. Moreover, the depletion was circular and had a maximum spatial extent of approximately $10^{\circ}$ in latitude and $10^{\circ}$ in longitude. However, contrary to the simulation, there is little evidence of the depletion after 21:30 UT. Moreover, the GNSS data indicated a maximum TEC decrease of approximately $4 \, \mathrm{ TECU}$ compared with the approximately $10.8 \, \mathrm{ TECU}$ decrease found in the simulation.


\begin{figure}
\includegraphics[width=\linewidth]{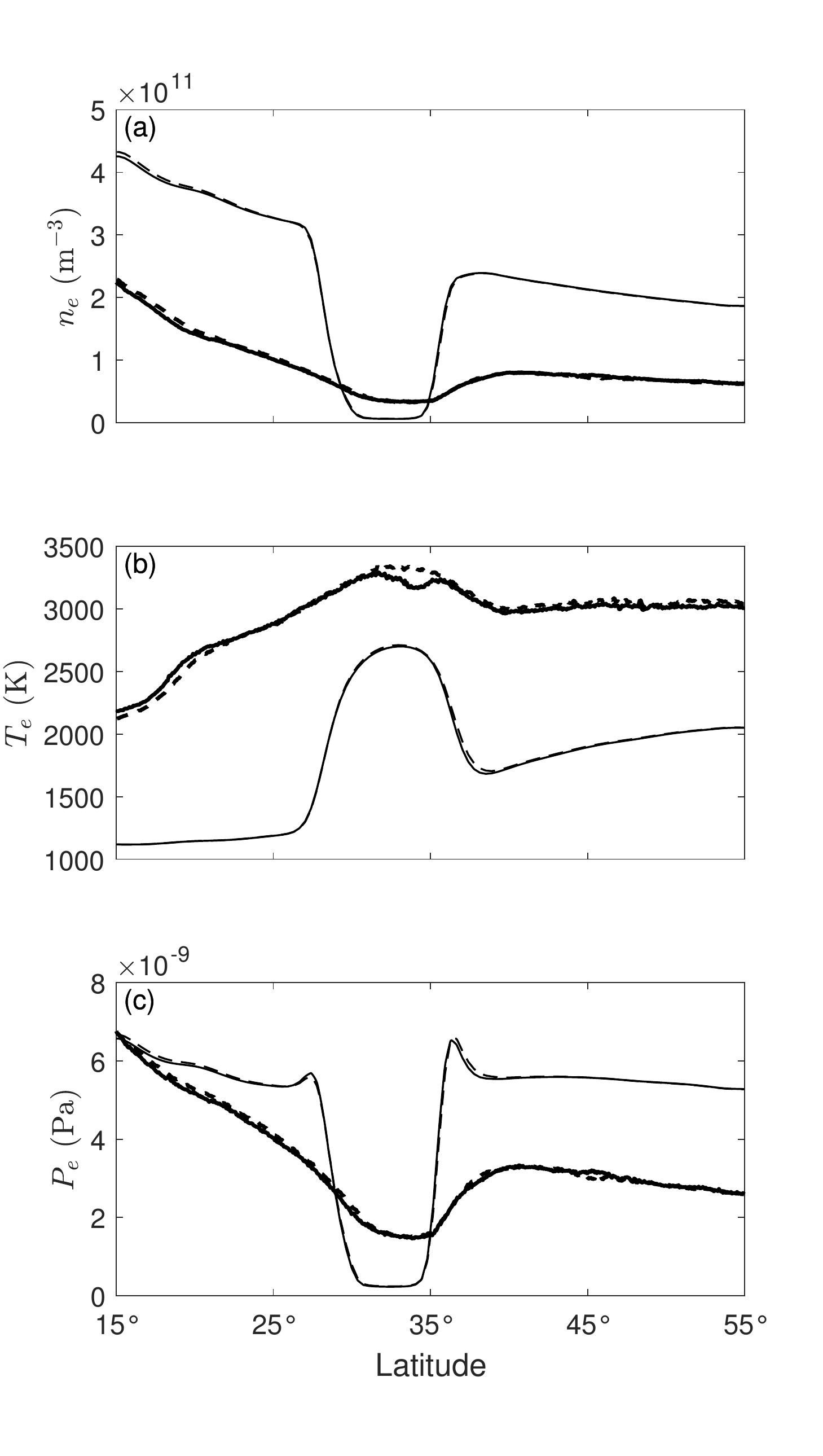}
  \caption{Electron (a) number density, (b) temperature, and (c) pressure measured by the Swarm-A (thick solid line) and Swarm-C (thick dashed line) satellites following the FORMOSAT-5 launch. Corresponding predictions from the GITM simulation assuming complete combustion at 19:50 UT are also shown (thin lines).}
  \label{fig_F5_SWARM_data}
\end{figure}

The ionospheric depletion due to the FORMOSAT-5 launch was also evident in Swarm Langmuir probe data \cite{jin2019} for 24 August 2017, plotted in Fig.~\ref{fig_F5_SWARM_data}. Park \cite{park2016} performed a similar analysis on both Swarm and DMSP Langmuir probe data and argued that observed depletions in the absence of geomagnetic, seismic, or tropospheric disturbances were due to rocket exhaust rather than natural phenomena. Local minima in $n_e$ are encountered by Swarm-A and Swarm-C at $33^{\circ}11' 30'' \, \mathrm{N}$, $123^{\circ}12' 27'' \, \mathrm{W}$ and $33^{\circ}37' 03'' \, \mathrm{N}$, $121^{\circ}45' 39'' \, \mathrm{W}$ respectively. Such minima were not seen in data for 23 August or 25 August. In each case the minima corresponded to an approximate altitude of $442 \, \mathrm{km}$ and time 19:48:50 UT. As the signal in $n_e$ is similar for the two satellites, the midpoint between the aforementioned locations can be assumed to approximate the centre of the depletion. This would imply that the minimum travelled at a mean ground track velocity of $66 \, \mathrm{m.s}^{-1}$ at $231^{\circ} \, \mathrm{N}$ from the launch site. This location corresponds closely to the TEC minimum in the GITM simulation at 19:50 UT, estimated to lie at $33^{\circ}12' 04'' \, \mathrm{N}$, $122^{\circ}36' 27'' \, \mathrm{W}$ and implying a ground track velocity of $72 \, \mathrm{m.s}^{-1}$ at $228^{\circ} \, \mathrm{N}$.

\begin{figure}
\centering
\subfloat[GITM simulation]{
\includegraphics[width=\linewidth]{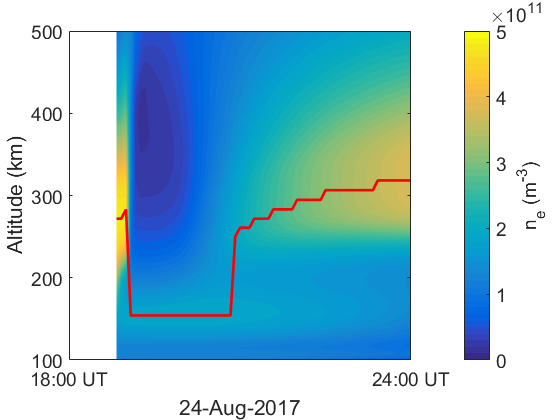}
\label{fig_PA836_ionograms_GITM}
}
\hspace{0mm}
\subfloat[Ionosonde data]{
\includegraphics[width=\linewidth]{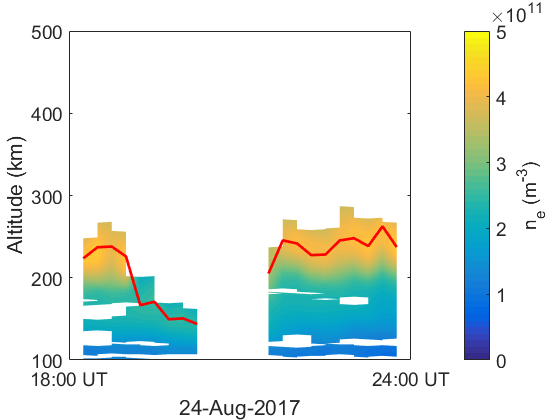}
\label{fig_PA836_ionograms_measured}
}
\caption{Electron number density profile and peak altitude (red line) at Point Arguello ionosonde location ($33^{\circ}36' \, \mathrm{N}$, $120^{\circ}36' \, \mathrm{W}$) following the FORMOSAT-5 launch.}
\label{fig_PA836_ionograms}
\end{figure}

The Swarm Langmuir probe observations may be compared with values interpolated from the GITM simulations at 19:50 UT, which are also plotted in Fig.~\ref{fig_F5_SWARM_data}. Ionospheric depletions are seen in both cases over a similar range of latitudes, approximately $33^{\circ} \pm 6^{\circ}$. However, the decrease in $n_e$ and increase in $T_e$ are far more pronounced in the GITM simulation than the Langmuir probe data. This difference implies that the quantity of rocket exhaust introduced into the upper thermosphere was overestimated in the numerical model. It is also noted that the undisturbed $n_e$ and $T_e$ obtained from the simulation are respectively much higher and lower than the Langmuir probe data. The former discrepancy appears to reflect the overestimation of the height of the F-region peak by GITM and the IRI model which initialises it, which can be seen in Fig.~\ref{fig_PA836_ionograms}.

Finally, the ionospheric depletion due to the FORMOSAT-5 launch was evident in vertical incidence soundings taken by the Point Arguello ionosonde. These measurements, plotted in Fig.~\ref{fig_PA836_ionograms_measured}, show rapid decreases in the maximum electron number density and altitude at which it occurred following the launch which are qualitatively similar to those found using the GITM model, plotted in Fig.~\ref{fig_PA836_ionograms_GITM}. Both the observational data and simulation indicated large depletions of the F-region with little impact on lower regions of the ionosphere. By 21:30 UT, the F-region had partly recovered in each case, and the electron number density maximum altitude had returned to near its value prior to launch. However, the ionosonde measurements show that the F-region peak prior to the launch was at lower altitude than in the IRI model which is used to intialise GITM. Furthermore, following the recovery of the F-region the ionosonde measurments show that this peak was lower than in the GITM simulation. Consequently, it is believed that GITM underestimated F-region electron production rates, which decrease with altitude. This is thought to have caused the model to underestimate the rate at which the ionosphere recovered from the depletion.

In contrast, the aforementioned features were not clearly observed in ground-based GNSS, satellite Langmuir probe, or ionosonde data for 17 January 2016 following the Jason-3 launch. This was unsurprising, given the much smaller and shorter-lived ionospheric depletion predicted by the GITM simulation. Moreover, ground-based GPS TEC and satellite Langmuir probe $n_e$ data following the Jason-5 launch had greater magnitude and greater spatio-temporal variation than those following the FORMOSAT-5 launch, making it more difficult to detect ionospheric depletion. Data from the Point Arguello ionosonde were sparse for the relevant period.

\section{Discussion} \label{sec_discussion}
\subsection{Comparison of launches}


Both the GITM simulations and observational data show that the FORMOSAT-5 launch resulted in a far larger and longer-lived depletion of the ionosphere than the Jason-3 launch, with several factors contributing to this difference. As noted above, the steeper trajectory in the former case resulted in a far greater proportion of exhaust gases being deposited at heights coinciding with the F-region and remaining in this region longer before falling to lower layers of the atmosphere. Additionally, the mass flow rate from the second stage in the latter case was $14 \%$ larger than in the former case. Moreover, UV flux was lesser in the FORMOSAT-5 launch case resulting in a lesser electron production rate. The F-region peak was higher in this case, as seen in Fig.~\ref{fig_altitude}, corresponding to a lower electron production rate.

Due to the different trajectories of the Jason-3 and FORMOSAT-5 launches, the approximation of the rocket exhaust gasses as expanding diffusively from a point source at rest would have affected the corresponding simulated results differently. Values of $L$ in the upper thermosphere in Fig.~\ref{fig_length_scale} are significant compared with the dimensions of the ionospheric depletions being considered. This comparison suggests that advection from the point of release is important in determining the spatial distribution of the rocket exhaust products. In the case of the FORMOSAT-5 launch, the near vertical trajectory would result in rocket exhaust gasses coming to rest at significantly lower altitudes than modelled here. The simulation would therefore in this case be expected to overestimate the overall ionospheric depletion, $\Delta N_e$, and consequently local reductions in TEC, as well as the longevity of these effects. In contrast, during the near horizontal phase of the trajectory for the Jason-3 launch, the exhaust gasses would have travelled significant horizontal distances. Thus, the simulation would be expected to overestimate the concentrations of rocket exhaust gasses and therefore local reductions in TEC.

\subsection{Comparison of simulation and observational data}


The size, shape, and initial motion of the ionospheric depletion in the first $2$ hours following the FORMOSAT-5 launch were modelled reasonably accurately by the GITM simulation. According to the previous analysis of Chou et al. \cite{chou2018} the maximum spatial extent was $6.36 \times 10^5 \, \mathrm{km^2}$, within $15\%$ of the simulated value. This suggests that the diffusion of the exhaust gasses through the thermosphere was modelled well by GITM. The horizontal movement of the ionospheric depletion was largely determined by that of the rocket exhaust gasses and therefore the horizontal winds in the thermosphere. Thus, the close agreement between simulation and observation regarding the location of the depletion during the Swarm satellite passes implies that horizontal winds were predicted accurately by GITM, which was initialised based on the empirical HWM07 model \cite{drob2008}.

In contrast, there were significant discrepancies between the simulations and observations regarding TEC and electron number density depletions and their longevity. These were likely due in large part to approximations made in the rocket exhaust source model. As noted above, the initial downward velocity of rocket exhaust molecules would result in their reaching lower altitudes before entering a diffusive expansion regime. This would reduce their impact on electron concentrations in the upper F-region, which takes longer to recover from depletions. Moreover, it results in the exhaust molecules falling below the F-region sooner and being depleted through charge exchange reactions more rapidly. The initial horizontal velocity of rocket exhaust molecules would disperse them over a wider area before entering the diffusive expansion regime. This would reduce the magnitude of TEC and electron number density depletions. The simulations described above neglected condensation of rocket exhaust, which reduces the number of molecules available for charge exchange reactions. In a previous study, condensation reduced the amount of \ch{H_2O} by $16.7 \%$ \cite{mendillo1981}.

Differences between the simulated and observed depletions also resulted from the ionosphere and thermosphere models used in this work. The IRI and GITM models provided reasonable estimates for the background TEC in both of the cases examined, as seen by comparison of the TEC maps Fig.~\ref{fig_GITM_TEC_map_data} and Fig.~\ref{fig_MAPGPS_TEC_data}. However, these models overestimated the height at which the F-region peak occurred without the depletion, as shown in Fig.~\ref{fig_PA836_ionograms}. As electron production rates decrease with altitude, this disagreement in turn contributes to the overestimate of the recovery time by the simulation. Furthermore, the boundary conditions of the regional GITM simulation do not incorporate general global circulation patterns. Thus, the accuracy of the simulation is expected to degrade significantly with time, particularly post sunset when day-to-night transport is neglected.

The GNSS TEC data were derived based on ionospheric pierce points at $450 \, \mathrm{km}$ altitude \cite{rideout2006}. However, the numerical modelling and ionosonde data shown in Fig.~\ref{fig_PA836_ionograms} indicate that the ionospheric plasma is predominantly concentrated at lower altitudes, particularly within the region affected by the depletion. Thus the depletion seen in Fig.~\ref{subfig_MAPGPS_F5_TEC_data} may be somewhat distorted. The oblique angles at which the GNSS measurements were taken together with the vertical distribution of the ionosphere limited resolution in latitude and longitude. This issue could have helped obscure the narrow ionospheric depletion the simulation suggested would occur following the Jason-3 launch.

Prior studies have indicated lifetimes of ionospheric depletions due to rocket launches ranging from $0.5$ to $6$ hours \cite{park2016}. This range is consistent with observational data for the FORMOSAT-5 launch case and the simulation of the Jason-3 launch case, but not the simulation of the FORMOSAT-5 launch case. However, large rocket launches have typically occurred over sea, making it difficult to take observations of ionospheric depletions throughout their evolution. Therefore, the possibility of rocket launches causing very long lived depletions by depositing exhaust gasses high in the ionosphere, as seen in the FORMOSAT-5 simulation, cannot be precluded.

\subsection{Future work}





Further simulations using the modified version of GITM could be used to investigate the sensitivity of ionospheric depletions to launch trajectory, the upper atmosphere environment, and physical and chemical parameters. The present work indicated the strong dependence of the ionospheric depletion on the altitude reached by the second stage and this should be further investigated. It would be possible to use the numerical model to consider the effects of increased solar and geomagnetic activity on the formation and recovery of ionospheric depletions. Variation with the local time and latitude of the launch might also be considered. Uncertainties in ionospheric depletion behaviour due to those in reaction rate and diffusion coefficients may be quantified through a sensitivity analysis.

An improved model of the rocket exhaust gas source could be obtained through the direct simulation Monte-Carlo (DSMC) method. This technique was previously used to model the transport of rocket exhaust in order in order to study ionospheric interactions of rocket exhaust by Bernhardt et al. \cite{bernhardt2012}. Rocket exhaust plume gasses push aside ionospheric plasma, resulting in a redistribution of $n_e$ on much shorter time-scales than the ionospheric depletions considered above. This is referred to as the snowplow effect and could be incorporated into such a DSMC simulation. The DSMC technique developed by Zhong et al. \cite{zhong2005} could also be used to determine the amount of exhaust lost through condensation.

It is envisaged that GITM could be further modified in future to incorporate other ionospheric effects of rocket launches. On shorter time-scales than considered above, small amplitude wave disturbances were observed propagating away from the rocket trajectory. These were generated by the expansion of the rocket exhaust gasses subsequent to their addition to the simulation (after $300 \, \mathrm{s}$ analytical expansion using eq.~\ref{gas_diffusion_point}), which did not include the sub-grid scale dynamics responsible for the waves which drive observed travelling ionospheric disturbances accompanying rocket launches. However, appropriate source terms could be incorporated into GITM to represent the excitation of the acoustic and atmospheric gravity waves during the passage of rockets through the thermosphere. The model described above may be used to study the potential for the rocket launches to trigger equatorial plasma bubbles. Using GITM to comprehensively model the state of the ionosphere-thermosphere system would enable the study of this phenomenon by coupling to an ionospheric code which self-consistently solves for electric fields, such as SAMI3 \cite{huba2008}.


\section{Conclusions}
A modified version of GITM has been developed to model ionospheric depletions due to rocket launches. The rocket exhaust gasses \ch{H_2O}, \ch{H_2}, and \ch{CO_2} were implemented as major species. A source term was added incorporating an initial diffusive expansion which was modelled analytically. Combined charge exchange and dissociative recombination reactions of these species were incorporated into the GITM chemical model.

The numerical model was applied to the Jason-3 and FORMOSAT-5 launches. Magnitude and longevity of the resulting depletions were found to depend on the altitudinal distribution of the rocket exhaust species which determined their residence times in the F-region. Thus, it was found that the steeper trajectory in the FORMOSAT-5 launch case produced larger and longer-lived depletions. Comparison of results for plume compositions determined based on complete combustion and CEA output indicated that the ionospheric depletion was not sensitive to this composition.

Results of the FORMOSAT-5 launch simulation were compared with GNSS, ionosonde, and satellite Langmuir probe measurements. The horizontal movement and expansion of the depletion following the FORMOSAT-5 launch in the numerical model were in reasonable agreement with GNSS TEC and Swarm Langmuir probe electron number density observations. However, the simulated depletions of TEC and electron number density and the time period over which they occurred were much greater than those observed. These disagreements were ascribed primarily to the neglect of the downward transport of rocket exhaust gasses prior to diffusive expansion and the overestimation of the F-region peak height by IRI and GITM.

\section{Acknowledgments}
This work was carried out with funding from the Royal Australian Air Force. Resources from the National Computational Infrastructure (NCI), supported by the Australian Government were used in this work. We thank Aaron Ridley for his invaluable guidance regarding the GITM code he has developed. Additionally, we gratefully acknowledge James Gilmore for providing data format conversion tools used in this work.

We acknowledge several organisations for supplying empirical ionospheric data. MAPGPS TEC map data were obtained through the Madrigal CEDAR Database, accessible via the World Wide Web at http://cedar.openmadrigal.org/index.html/. This service is provided by the Massachusetts Institute of Technology under support from US National Science Foundation grant AGS-1242204. Langmuir probe data from the ESA Swarm spacecraft were derived from the SWARM Data Access website, which can be found at https://swarm-diss.eo.esa.int/. NOAA ionosonde data were also used, which can be downloaded via FTP from ftp://ftp.ngdc.noaa.gov/ionosonde.


\begin{thebibliography}{10}

\bibitem{booker1961}
H.~Booker.
\newblock A local reduction of {F}-region ionization due to missile transit.
\newblock {\em Journal of Geophysical Research}, 66(4):1073--1079, 1961.

\bibitem{mendillo1975}
M.~Mendillo, G.~S. Hawkins, and J.~A. Klobuchar.
\newblock A sudden vanishing of the ionospheric {F} region due to the launch of
  skylab.
\newblock {\em Journal of Geophysical Research}, 80(16):2217--2228, 1975.

\bibitem{mendillo1982}
M.~Mendillo and J.~Baumgardner.
\newblock Optical signature of an ionospheric hole.
\newblock {\em Geophysical Research Letters}, 9(3):215--218, 1982.

\bibitem{wand1984}
R.~H. Wand and M.~Mendillo.
\newblock Incoherent scatter observations of an artificially modified
  ionosphere.
\newblock {\em Journal of Geophysical Research: Space Physics},
  89(A1):203--215, 1984.

\bibitem{park2016}
J.~Park, H.~Kil, C.~Stolle, H.~L{\"{u}}hr, W.~R. Coley, A.~Coster, and Y.~S.
  Kwak.
\newblock Daytime midlatitude plasma depletions observed by {S}warm: Topside
  signatures of the rocket exhaust.
\newblock {\em Geophysical Research Letters}, 43(5):1802--1809, 2016.

\bibitem{furuya2008}
T.~Furuya and K.~Heki.
\newblock Ionospheric hole behind an ascending rocket observed with a dense
  {GPS} array.
\newblock {\em Earth, Planets and Space}, 60(3):235--239, 2008.

\bibitem{mendillo2008}
M.~Mendillo, S.~Smith, A.~Coster, P.~Erickson, J.~Baumgardner, and C.~Martinis.
\newblock Man-made space weather.
\newblock {\em Space Weather}, 6(90):215--218, 2008.

\bibitem{chou2018}
M.-Y. Chou, M.-H. Shen, C.~C.~H. Lin, J.~Yue, C.-H. Chen, J.-Y. Liu, and J.-T.
  Lin.
\newblock Gigantic circular shock acoustic waves in the ionosphere triggered by
  the launch of {FORMOSAT}-5 satellite.
\newblock {\em Space Weather}, 16(2):172--184, 2018.

\bibitem{bernhardt1975}
P.~A. Bernhardt, C.~G. Park, and P.~M. Banks.
\newblock Depletion of the {F}2 region ionosphere and the protonosphere by the
  release of molecular hydrogen.
\newblock {\em Geophysical Research Letters}, 2(8):341--344, 1975.

\bibitem{bernhardt1979}
P.~A. Bernhardt.
\newblock Three-dimensional, time-dependent modeling of neutral gas diffusion
  in a nonuniform, chemically reactive atmosphere.
\newblock {\em Journal of Geophysical Research}, 84(A3):793--802, 1979.

\bibitem{mendillo1978}
M.~Mendillo and J.~M. Forbes.
\newblock Artificially created holes in the ionosphere.
\newblock {\em Journal of Geophysical Research: Space Physics},
  83(A1):151--163, 1978.

\bibitem{ridley2006}
A.~J. Ridley, Y.~Deng, and G.~T{\'{o}}th.
\newblock The global ionosphere-thermosphere model.
\newblock {\em Journal of Atmospheric and Solar-Terrestrial Physics},
  68(8):839--864, 2006.

\bibitem{bilitza1990}
D.~Bilitza, K.~Rawer, L.~Bossy, I.~Kutiev, K.~I. Oyama, R.~Leitinger, and
  E.~Kazimirovsky.
\newblock International reference ionosphere 1990.
\newblock Technical report, NASA Science Applications Research, Lanham, 1990.

\bibitem{picone2002}
J.~M. Picone, A.~E. Hedin, D.~P. Drob, and A.~C. Aikin.
\newblock {NRLMSISE-00} empirical model of the atmosphere: Statistical
  comparisons and scientific issues.
\newblock {\em Journal of Geophysical Research: Space Physics},
  107(A12):SIA--15, 2002.

\bibitem{mendillo1981}
M.~Mendillo.
\newblock The effect of rocket launches on the ionosphere.
\newblock {\em Advances in Space Research}, 1(2):275--290, 1981.

\bibitem{anicich2003}
V.~G. Anicich.
\newblock An index of the literature for bimolecular gas phase cation-molecule
  reaction kinetics.
\newblock Technical report, JPL, Pasadena, 2003.

\bibitem{lindinger1974}
W.~Lindinger, F.~C. Fehsenfeld, A.~L. Schmeltekopf, and E.~E. Ferguson.
\newblock Temperature dependence of some ionospheric ion-neutral reactions from
  {$300^{\circ}-900^{\circ}$} {K}.
\newblock {\em Journal of Geophysical Research}, 79(31):4753--4756, 1974.

\bibitem{smith1978}
D.~Smith, N.~G. Adams, and T.~M. Miller.
\newblock A laboratory study of the reactions of {$\text{N}^+$},
  {$\text{N}_2^+$}, {$\text{N}_3^+$}, {$\text{N}_4^+$}, {$\text{O}^+$},
  {$\text{O}_2^+$}, and {$\text{NO}^+$} ions with several molecules at 300 {K}.
\newblock {\em The Journal of Chemical Physics}, 69(1):308--318, 1978.

\bibitem{vejbychristensen1997}
L.~Vejby-Christensen, L.~H. Andersen, O.~Heber, D.~Kella, H.~B. Pedersen, H.~T.
  Schmidt, and D.~Zajfman.
\newblock Complete branching ratios for the dissociative recombination of
  {$\text{H}_2 \text{O}^+$}, {$\text{H}_3 \text{O}^+$}, and {$\text{CH}_3^+$}.
\newblock {\em The Astrophysical Journal}, 483(1):531--540, 1997.

\bibitem{schunk2009}
R.~W. Schunk and A.~F. Nagy.
\newblock {\em Ionospheres: physics, plasma physics, and chemistry}.
\newblock Cambridge University Press, Cambridge, second edition edition, 2009.

\bibitem{fuller1966}
E.~Fuller, P.~Schettler, and J.~Giddings.
\newblock A new method for prediction of binary gas-phase diffusion
  coefficients.
\newblock {\em Industrial and Engineering Chemistry}, 58(5):18--27, 1966.

\bibitem{bernhardt1979b}
P.~A. Bernhardt.
\newblock High-altitudegas releases: Transition from collisionless flow to
  diffusive flow in a nonuniform atmosphere.
\newblock {\em Journal of Geophysical Research}, 84(A8):4341--4354, 1979.

\bibitem{jarvinen1966}
P.~O. Jarvinen, J.~A.~F. Hill, J.~S. Draper, and R.~E. Good.
\newblock High altitude rocket plumes.
\newblock Technical report, MITHRAS, Cambridge, 1966.

\bibitem{gordon1996}
S.~Gordon and B.~J. McBride.
\newblock Computer program for calculation of complex chemical equilibrium
  compositions and applications.
\newblock Technical report, NASA, Lewis Research Center, Cleveland, 1996.

\bibitem{rideout2006}
W.~Rideout and A.~Coster.
\newblock Automated {GPS} processing for global total electron content data.
\newblock {\em GPS Solutions}, 10(3):219--228, 2006.

\bibitem{yizengaw2008}
E.~Yizengaw, M.~B. Moldwin, D.~Galvan, B.~A. Iijima, A.~Komjathy, and A.~J.
  Mannucci.
\newblock Global plasmaspheric {TEC} and its relative contribution to {GPS}
  {TEC}.
\newblock {\em Journal of Atmospheric and Solar-Terrestrial Physics},
  70(11-12):1541--1548, 2008.

\bibitem{jin2019}
Y.~Jin, A.~Spicher, C.~Xiong, L.~B. Clausen, G.~Kervalishvili, C.~Stolle, and
  W.~J. Miloch.
\newblock Ionospheric plasma irregularities characterized by the swarm
  satellites: Statistics at high latitudes.
\newblock {\em Journal of Geophysical Research: Space Physics},
  124(2):1262--1282, 2019.

\bibitem{drob2008}
D.~P. Drob, J.~T. Emmert, G.~Crowley, J.~M. Picone, G.G. Shepherd, W.~Skinner,
  P.~Hays, R.J. Niciejewski, M.~Larsen, C.Y. She, and J.W. Meriwether.
\newblock An empirical model of the earth's horizontal wind fields: {HWM07}.
\newblock {\em Journal of Geophysical Research}, 113(A12), 2008.

\bibitem{bernhardt2012}
P.~A. Bernhardt, J.~O. Ballenthin, J.~L. Baumgardner, A.~Bhatt, I.~D. Boyd,
  J.~M. Burt, R.~G. Caton, A.~Coster, P.~J. Erickson, J.~D. Huba, and G.~D.
  Earle.
\newblock Ground and space-based measurement of rocket engine burns in the
  ionosphere.
\newblock {\em IEEE transactions on plasma science}, 40(5):1267--1286, 2012.

\bibitem{zhong2005}
J.~Zhong, M.~I. Zeifman, and D.~A. Levin.
\newblock Modeling of argon condensation in free-jet expansions with the
  {DSMC} method.
\newblock {\em AIP Conference Proceedings}, 762(1):391--395, 2005.

\bibitem{huba2008}
J.~D. Huba, G.~Joyce, and J.~Krall.
\newblock Three-dimensional equatorial spread {F} modeling.
\newblock {\em Geophysical Research Letters}, 35(10), 2008.

\end{thebibliography}
\end{document}